# Mapping reaction mechanism during overcharge of a LiNiO2/Graphite-silicon lithium-ion battery: a correlative operando approach by simultaneous gas analysis and synchrotron scattering techniques

Quentin Jacquet[1#*], Irina Profatilova[2#], Loïc Baggetto[2], Bouthayna Alrifai[2], Elisabeth Addes[2], Paul Chassagne[2], Nils Blanc[3], Samuel Tardif[4], Lise Daniel[2], Sandrine Lyonnard[1*]

[1] Univ. Grenoble Alpes, CEA, CNRS, Grenoble INP, IRIG, SyMMES, 38000 Grenoble, France.
[2] Univ. Grenoble Alpes, CEA, Liten, DEHT, 38000 Grenoble, France
[3] European Synchrotron Radiation Facility (ESRF), CS 40220, 71 Avenue des Martyrs, 38043 Grenoble, France
[4] University Grenoble Alpes, CEA, CNRS, IRIG, MEM, 38000 Grenoble, France.

*Corresponding: quentin.jacquet@cea.fr, sandrine.lyonnard@cea.fr
# equal contribution

Abstract :
Li-ion battery degradation processes are multi-scale, heterogeneous, dynamic and involve multiple cell components through cross talk mechanisms. Correlated operando characterization capable of measuring several key parameters are needed to accelerate understanding on these complex degradation processes. In particular, degradation mechanisms during overcharge of LiNiO$_2$/Graphite-Silicon is well known at the material level featuring O$_2$ gas release and concomitant surface reconstruction of LiNiO$_2$. However, there are still debates regarding the role of high voltage O1 phase formation on gas production and no information on the effect of produced gases on the cell components (anode or sensors), or effect of overcharge on electrode level behavior. In this work, we simultaneously measured the gas produced using operando mass spectrometry while spatially resolving nanostructure and lattice changes using operando micro SAXS/WAXS mapping during the formation and over charge of a LiNiO$_2$/Gr-Si pouch cell. This new correlated operando characterization experiment allowed to (1) confirm the absence of O1 phase even with substantial gas produced at end of charge, (2) unveil the effect of gases on reference and negative electrodes, (3) show that overcharge increases in-plane reaction heterogeneities by creating local degraded spots lagging behind the ensemble electrochemistry. These findings will be important to optimize ageing of devices based on similar chemistries, in particular Ni-rich NMC, while showing the strength of correlated characterization leading to more efficient and robust information on complex mechanisms.

**Introduction**

   The widespread development of electrical vehicles needed for the electrification of the transportation sector will require mining and refining very large quantities of raw materials, some of these being scarce or geographically concentrated[1,2]. Increasing the lifetime and the energy densities of lithium-ion batteries (LIBs) is therefore crucial to reduce the number of batteries that need to be produced and recycled, as well as to extend the driving range and minimize the environmental impact of the other materials (packaging, engine, wiring, etc.).

LiNiO$_2$ (LNO) and Graphite-Silicon composites (Gr-Si) are promising materials for high-energy density Li-ion batteries (LIBs)[3]. LNO is the end member of the Ni-rich layered oxide family of general composition LiNi$_x$Mn$_y$Co$_z$O$_2$ so called NMC (with x + y + z = 1 and x ≥ 0.6). These Ni-rich materials offer the dual benefit of high energy density and low Co content but suffer from higher degradation upon cycling. In particular, (1) lattice oxygen is partially released at high voltage causing surface reconstruction, electrolyte decomposition and gas evolution leading to battery safety risks[4] and (2) severe intra-particle strains, arising from large unit cell parameter changes during cycling, give rise to particle fracture creating disconnected particles and newly exposed surfaces to the electrolyte[5]. On the anode side, graphite has been used for decades in Li-ion batteries, and Gr-Si composite, which features a higher energy density thanks to the alloying reaction between Li and Si, also suffers from poorer cycling stability. This originates partly in failure of the *in situ* formed passivating film (SEI, solid electrolyte interphase) upon repeated volume expansion/shrinkage of Si particles, leading to constant electrolyte consumption at the anode surface[6]. Anode-side and cathode-side degradations cannot only be studied separately as decomposition products formed at one electrode often lead to gaseous or dissolved species which migrate in the battery and react with other cell components, process commonly referred to as "crosstalk"[7]. Moreover, LIB degradation processes can depend on active material position in the cell and local defects. Indeed, it is known that electrode edges and/or active material areas around defects can have different states of charge (SoC) or reaction kinetics compared to the average electrode reaction, hence different degradation mechanisms[8]. Finally, degradation depends also on battery usage, with fast charging or overcharge being particularly harsh conditions[9]. Clearly, degradation mechanisms in Li-ion batteries, and in particular batteries containing LNO/Gr-Si, are very complex because numerous, dynamic, occurring over multiple length scales and involving multiple components[10]. While a consensus on specific degradation mechanisms start to emerge at the material level, as described earlier for Ni-rich materials or Gr-Si composites, much work is still needed to (1) spatially resolve degradation processes at the material and cell level, and to (2) understand the correlation between different degradation mechanisms. For example, it is important to unravel interdependencies between cathode/anode phenomena, interactions between gaseous/soluble/solid species, or distribution of active/defective zones in batteries in nominal and abusive conditions.

Along that line, we focus on understanding the correlations between structural evolution of LNO/Gr-Si materials and gas release especially during the formation cycle and overcharge at 5 V vs. Li/Li$^+$. Moreover, we evaluate the spatial homogeneity of the cell degradation at high voltage to link the observed effects to the cell geometry. To address these questions, *operando* multi-probe characterization is necessary[11] and can be done according to two general strategies being (1) carrying out several individual characterizations to measure specific parameters or (2) developing new instruments/techniques capable of measuring simultaneously as many key parameters as possible [12,13]. While the first approach is easier to execute, it still holds many challenges regarding for example the sample transfer (making sure that the sample remains intact between two measurements), or the use of the same test conditions and correlation of datasets (biases introduced by samples and measurements statistics). The second strategy is most desired because one gets information on the same sample, in the same cell and at the same time but requires developing instruments and *operando* electrochemical cells accommodating several probes. Focusing on the correlation between degassing and structural evolution of the active material, De Biasi *et al.* applied the first approach and measured averaged structural evolution in LNO/Li systems up to 4.5 V using *operando* laboratory X-ray diffraction (XRD) on pouch cell and gas formation measured with Online Electrochemical Mass Spectrometry (OEMS) using a different specially designed cell [14]. In terms of gas release, they observed the formation of $CO_2$ at 4.1 V vs Li/Li$^+$ corresponding to the onset of the H2 → H3 transition in LNO while oxygen is mostly produced after this transition. Other reports mentioned $CO_2$ onset at 3.8 V or 4.4 V with or without the presence of $O_2$ [15,16]. Structure wise, De Biasi *et al.* observed the classical sequence of H1 → M → H2 → H3 phase transitions in general agreement with literature[17,18]. However, there are still debates on the formation mechanism of the O1 phase observed in some reports and suggested to be related to $O_2$ loss[19,20]. To our knowledge, there is no report of simultaneous *operando* gas measurements and structural evolution of active materials during cycling (same sample, same cell, same time) in LIBs. In addition, to the best of our knowledge, there is no information available in general regarding the relationship between overcharge and gassing on the cell level reaction heterogeneity. Note that resolving *operando* in-plane degradation in pouch cell during fast charge (locating Li plating in single layer pouch cell) and ageing (reaction heterogeneity in prismatic cells) has been reported for NMC/Gr-Si systems[21–23].

The present work reports the simultaneous measurements of gases and mapping of active material structural evolution in a single layer pouch cell during formation and overcharge cycles. The LNO/Gr-Si pouch cell is connected to a mass spectrometer and entirely scanned with a synchrotron X-ray microbeam during cycling to measure the time-resolved wide angle and small angle X-ray scattering (WAXS and SAXS) patterns. First, we obtain a direct correlation between the averaged structural evolution of the LNO and Gr materials determined by diffraction (WAXS) and the gas release. We find that $CO_2$ and $O_2$ are released at the end of the H2-H3 transition without noticeable bulk O1 phase formation in LNO. Second, we quantify the in-plane heterogeneities in both electrodes by spatially resolving the lattice parameter changes. The observed heterogeneities are found to originate from three main sources – (1) edge effects due to oversized Gr-Si electrode, (2) reference electrode and (3) the gases generated during overcharge. Indeed, the discharge after the overcharge features very local spots (smaller than the mm) lagging behind the overall electrochemistry. Some of these positions correspond to visible fabrication defects of the pouch cell. We propose that gas bubbles formed during the overcharge gather and grow at cell defects inducing the observed heterogeneity.

**Experiment details**

**Pouch cell**: Single layer pouch cells are assembled using $LiNiO_2$ electrode (94%wt. active material, 3%wt. C65, 3%wt. PVDF - BASF) and Gr-Si electrode (85%wt of active material including 11%wt of Si nanoparticles - CIDETEC). Electrode capacity loadings are 3 mAh.cm$^{-2}$ and 3.3 mAh.cm$^{-2}$, respectively. The separator is two layers of DreamWeaver Gold, non-woven sheet made of Kevlar-type fibers, and the electrolyte is 1.3 M $LiPF_6$ in ethylene carbonate (EC) and 10 wt.% fluoroethylene carbonate (FEC). Organic carbonate solvents with high vapor pressures such as dimethyl carbonate (DMC) and ethyl methyl carbonate (EMC) were not used in this study to prevent the mass-spectrometer contamination. The separator was chosen to ensure good wettability with the electrolyte. The cell was assembled in a dry room with a dew point of -40°C and all the components were dried prior to assembly at 105°C in vacuum for 24h. The electrode area was 2.3 x 2.3 cm$^2$ for LNO and 2.7 x 2.7 cm$^2$ for Gr-Si while the separator was larger to prevent short circuit (4.0 x 4.0 cm). The reference electrode is partially delithiated $LiFePO_4$ (LFP) of ca. 0.2 x 0.2 cm size. It was electrochemically pre-delithiated to achieve a LFP/FP phase fraction of 1/1 to ensure stable electrochemical potential of 3.42 V vs. Li/Li$^+$. The reference electrode was inserted between the cathode and the anode and electrically insulated by a layer of separator on each side. 1/32'' diameter PEEK tubes were sealed in the pouch cell using a thermosealing polymer, and served as gas inlet and outlet for the cell. Positive Al and negative Cu tabs were also sealed in the pouch cell using thermosealing polymer. After assembly, the pouch cell was pressed between two glassy carbon windows (4 x 4 cm$^2$ and 0.5 mm thick) using stainless steel perforated plates and clips. The accessible area for SAXS/WAXS observations was 3 x 3 cm$^2$. A picture of the cell and a schematic view of the investigated electrode stack are presented in Figure S1. Electrochemical cycling of the pouch cell was performed using a SP-300 potentiostat (BioLogic) in galvanostatic and potentiostatic modes by controlling the cell voltage while measuring both positive and negative electrode potentials against the reference electrode. One formation cycle was performed at C/13 with cut-off voltages of 4.2 V and 2.5 V vs. Li/Li$^+$ with a potentiostatic hold at the end of charge at 4.2 V. The hold ending conditions were achieved by either limiting the current to 0.24 mA or the step time to 2h. The overcharge cycle was then performed at C/9 with cut-off voltages of 5.0 – 2.5 V and a 3 h hold at 5 V at the end of overcharge. Note that the hold was interrupted after 30 min for a short rest time in open circuit voltage (OCV) of 7 min. The temperature during the tests was 26°C measuring at a temperature sensor placed against of external casing of the pouch.

**Gas analysis**

The pouch cell was mounted on the BM02 beamline at the European Synchrotron Radiation Facility (ESRF, Grenoble, France). The OEMS technique was applied to analyse the gases released during the electrochemical test of the pouch cell. For that purpose, a gas analysis line containing mostly 1/8'' metallic tubes was built-up around the cell. Ultra pure Ar BIP® (> 99.9999%) was used as a carrier gas. A gas purification column was installed before the electrochemical cell to remove any possible residual impurities from the tubing. A digital mass-flow controller was used to supply a constant flow of Ar of 2 mL.min$^{-1}$ to the cell. The gas analysis line was equipped with a vacuum pump and a bypass to be able to clean/purge the tubing prior to passing the gas through the cell. Flexible 1/32'' PEEK tubing sealed into the cell from both sides (Figure S1a) allowed a non-interrupted gas flow through the cell during its continuous movement. Hiden HPR20 S1000 triple filter quadrupole mass spectrometer with pulse ion counting electron multiplier detector was used for the measurements. The cell was purged with Ar for 3 h for stabilisation of baseline gas signals before starting the electrochemical test. The following mass-to-charge ratio (m/z) values were measured in Multiple Ion Detection mode: 44 ($CO_2$), 2 ($H_2$), 28 ($CO+C_2H_4$ mostly with minor contribution from $CO_2$), 29 ($CO + C_2H_4$ mostly), 32 ($O_2$), 15 ($CH_4$), 41 ($C_3H_6$). The volume of the electrochemical cell was 10 mL, giving a response time for OEMS analysis of 5 min. No calibration with a standard gas bottle was conducted during the experiment due to restricted beam time. However, the tests were repeated later in the laboratory for comparison with a gas quantification (Figure S10).
GC-MS measurements were conducted with an Agilent 7890A system using a J&W DB-200 122-2032 column. Injection occurred in split mode at 280 °C with 1 μL of sample carried by He gas with a 120 mL·min-1 flow rate. The column temperature ramp started at 40 ° until 200 °C at 10°C/min. The mass analyser was a MSD5975C (Agilent) with electron impact ionization.

**SAXS/WAXS mapping**

SAXS/WAXS mapping was performed on the French beamline BM02 at the European Synchrotron Radiation Facility (ESRF). A 65 x 100 microns beam at 18 keV was used ($\lambda$ = 0.6888011 Å). During the acquisition, the pouch cell was moved horizontally (x) and vertically (y) to scan the entire electrode surface, producing (x, y) maps of 31 x 31 pixels with pixel size of 1 mm in approx. 5 min. As the SAXS and WAXS modalities are available simultaneously by using two detectors, each pixel contained a WAXS and SAXS pattern. To reach this time resolution, "horizontal fly scans" were performed in which the pouch cell was continuously moved in the *x* direction from 0 to 31 mm while the shutter remained opened. Detector images were averaged over 1 mm pouch cell displacement. Therefore, each detector image was an average in the *x* direction of 1 mm of the sample. After every horizontal continuous scan, the pouch cell was displaced vertically by 1 mm (in *y* direction) and a new continuous scan in *x* was performed until the full 2D maps were produced. WAXS patterns were recorded using an imXPAD WOS detector, while the SAXS was recorded on an imXPAD S540 detector placed 11.6 and 3544 mm behind the sample, respectively. Sample-to-detector distance calibration was performed using reference materials: silver behenate ($AgC_{22}H_{43}O_2$) for SAXS and lanthanum hexaboride ($LaB_6$) for WAXS. Azimuthal integration was performed using PyFAI[24] and the patterns were normalized by $I_o$ (incoming photon flux) and $I_t$ (transmitted photon flux). Continuous acquisition was performed during the electrochemical cycling. Note that two detector shut-downs occurred during the formation cycle (middle of charge and of discharge) and provoked the absence of scattering data in these corresponding time lapse. After obtaining the WAXS patterns, data was re-calibrated to correct for the (small) difference in position between the $LaB_6$ reference and the electrode in the pouch cell. Calibration was performed pixel by pixel using Cu peak position as a reference. Note that no displacement of the pouch cell was observed during cycling. Background removal was performed using Prisma[25]. Moreover, peaks that did not evolve during the formation cycle and overcharge were considered as background and subtracted from the patterns. To get Li concentration in $LiNiO_2$, the centers of mass of (003) and (101) Bragg reflections were determined for every WAXS pattern and used to determine *a* and *c* lattice parameters. The Li concentration versus *a* lattice parameter calibration curve was fitted using charge of formation cycle and reference data[14], and used to convert *a* in Li concentration. To determine Li concentration in graphite electrode, the relative intensity and center of mass of peaks in the 1.91 – 1.838, 1.837 – 1.805, 1.8 – 1.761 and 1.73 – 1.76 Å$^{-1}$ regions, respectively corresponding to graphite, Stage 3, Stage 2/2L and Stage 1 ($LiC_6$) transitions were determined[26]. The Li concentration of each phase is determined based on the peak center of mass following the work of S. Tardif *et al.*[27] The overall Li concentration in graphite at every pixel position was obtained by averaging over the different phases weighted by their respective phase fractions calculated by the relative peak intensities. (111) Bragg reflection of crystalline silicon is observed in the WAXS pattern and used to qualitatively determine changes in crystalline Si amount. Radially-averaged 1D SAXS data mostly consist of a decaying intensity with a slope $Q^{-\alpha}$, without significant shape change during the electrochemical sequence, hence it was integrated between $4.10^{-3}$ and $3.10^{-2}$ Å$^{-1}$ to follow the averaged integrated intensity variations.

**Results**

**General principle of the experiment and beam damage assessment**

The experiment consists in imaging the active material (de)lithiation in *operando* conditions inside a single layer LNO/Gr-Si pouch cell while simultaneously following gas generation (Figure 1a). For that purpose, a 10 x 8 cm$^2$ pouch cell containing a 2.3 x 2.3 cm$^2$ positive electrode is connected to a mass spectrometer through a gas analysis line continuously purged with Ar carrier gas allowing the chemical analysis of gases produced. Simultaneously, the entire electrode stack is scanned using a 65 x 100 µm$^2$ X-ray beam mapping a zone of 3 x 3 cm$^2$ (Figure 1b). Wide and small angle scattering patterns are recorded simultaneously every 1 mm in the cell to resolve the atomic structure evolution of LNO and Gr with WAXS while also probing information on the nanoscale structure change using SAXS. These maps are recorded in 6 minutes, approx. 200 maps per cycle at C/13 and C/9, which corresponds to ca. 1% capacity change per map. A formation cycle up to 4.2 V and overcharge cycle up to 5 V were successively measured. Both conditions are known to produce gases due to SEI formation and decomposition of cathode material together with electrolyte oxidation at high voltage.

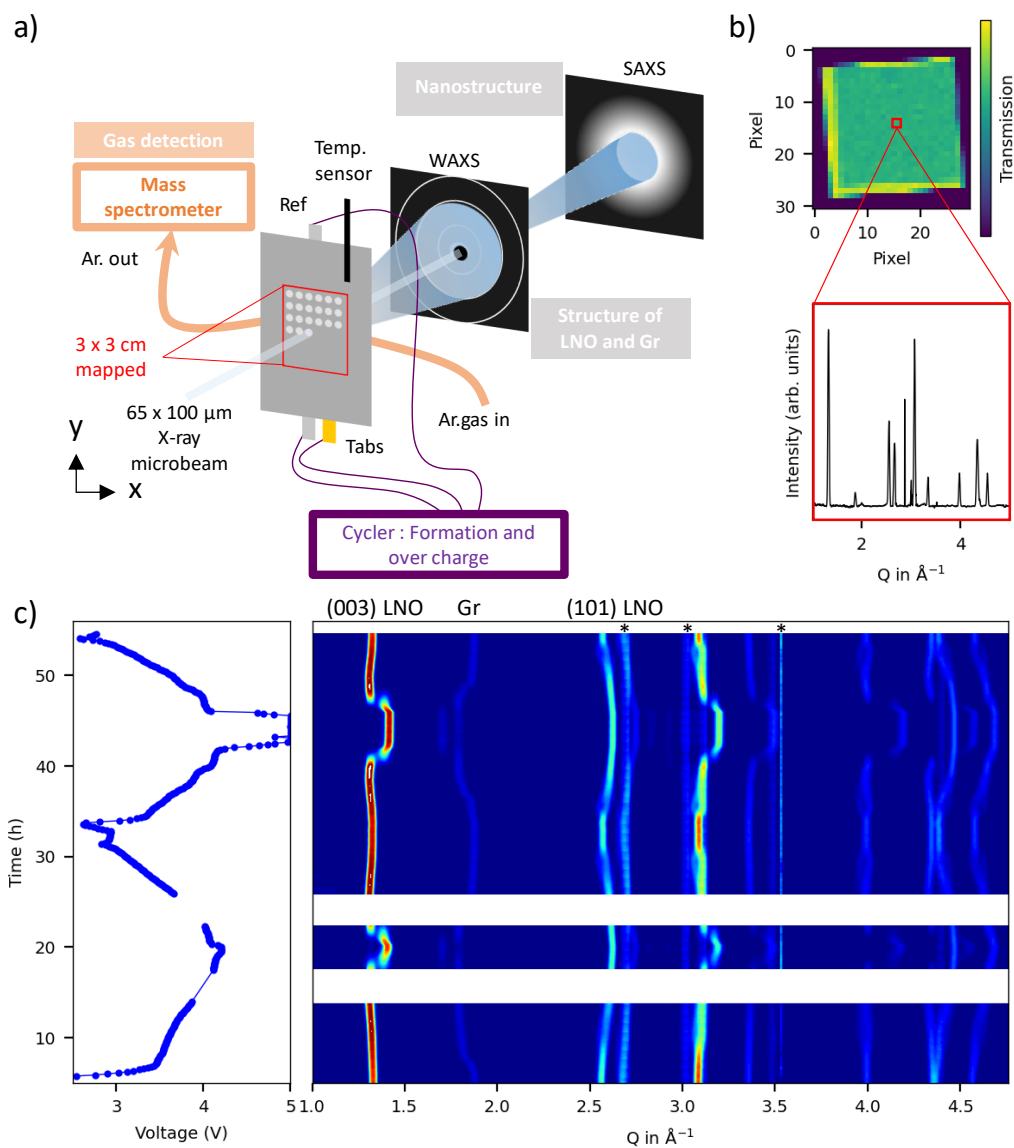

Figure 1: **Principle of the operando correlated SAXS/WAXS mapping & OEMS** a) schematic of the experimental set-up, b) 2D transmission map of the electrode stack – the center (green) region corresponds to the LNO/Gr-Si stack while the yellow borders are the region where only the oversized Gr-Si is present. Contours of the electrode stack appear dark because of stainless steel sample holder. Pixel size is 1*1 mm. Each pixel contains a full WAXS (1D zoom shown) and SAXS pattern. c) Electrochemical data during the formation and overcharge together with the 1D WAXS data averaged over the stack. Background and peaks from current collectors were removed during the data analysis. * mark the position of the current collector peak residuals while most of the peaks above 3 Å$^{-1}$ come from LNO. White stripes correspond to beam loss regions.

Charge and discharge voltage profiles for both cycles are composed of several plateaus corresponding to the various well-known transitions during (de)lithiation of LNO and Gr-Si electrodes (Figure 1c)[3,28]. Note that the electrochemical profile is almost identical to similar pouch cells measured in the lab – in absence of X-ray beam (Figure S2). The pouch cell is equipped with a LFP/FP reference electrode to access the individual voltage of both electrodes. When looking at the individual voltage of both electrodes recorded against the LFP/FP reference electrode, an important drop of the cathode and anode voltage during the 5 V (cell voltage) hold is observed for the cell tested at the beamline (Figure S2 c, e). In addition, subsequent to the 5 V hold, there is a drop of voltage of ca. 0.4 V at the beginning of the second discharge for both electrodes. The origin of this drop is nested in Li removal from the LFP of the reference electrode during the hold at 5 V, and is expected to increase the reference potential up to around 3.8 V vs. Li/Li$^+$. Removal of Li was quantified *operando* from the LFP/FP phase fraction obtained by WAXS and *ex situ* by *post-mortem* electrochemical analysis (Figure S3 and S4). The concentration change (quasi-complete delithiation of LFP into FP) led to the temporary increase of the reference electrode potential, which shows that the measured cathode and anode voltage drops are not due to actual change in electrochemical potential of active electrodes. Turning to the WAXS signal, diffraction peaks of LNO and Gr are clearly visible on the spatially averaged WAXS patterns shown in Figure 1c. The (111) Si peak at 2 Å$^{-1}$ is barely visible due to the small fraction of crystalline Si in the electrode (11 wt.%) and its nanometric size (30 nm) (Figure S5-6). For LNO, the peak around 1.3 Å$^{-1}$ corresponds to the (003) reflection of LNO (R-*3m*) and is indicative of the interlayer distance between the NiO$_2$ slabs of this layered material. It shifts towards lower Q values (larger interlayer distance) at the beginning of charge and higher Q values (smaller interlayer distances) at the end of charge with a strong shift at the very end corresponding to the formation of the so-called H3 phase[18]. Moving to the anode side, peaks in the region between 1.75 – 1.95 Å$^{-1}$ are indicative of the presence and interlayer distance of graphite and Li intercalated compounds. Typical reported phase sequence during lithiation features graphite (≈ 1.87 Å$^{-1}$), Stage X for X > 3 (≈ 1.85 Å$^{-1}$), Stage 3L (≈ 1.81 Å$^{-1}$), Stage 2/2L (≈ 1.79 Å$^{-1}$), and Stage 1 (≈ 1.70 Å$^{-1}$)[26]. The overall evolution of LNO and Gr diffraction peaks during charge and discharge follows similar trend to reported work[29]. The gas evolution profiles measured during cycling at the ESRF and in the lab (without X-ray exposure) using OEMS are compared Figure S7 and Figure S8. The data obtained in the lab feature a slightly better signal for some gases due to lower baselines thanks to a more efficient purge of the cell and line prior to electrochemical tests. Nonetheless, the gas evolution trends are essentially the same for both cells, confirming the absence of beam effect on gas production even at high voltage where the electrochemical decomposition of the electrolyte occurs.

**Formation cycle**

Having ruled out beam damage effect on electrochemistry, structural changes and gas signal, we move to a more quantitative analysis of the data starting with the formation cycle. Interlayer distance for Li$_x$C$_6$, *c* parameter for LNO and normalized SAXS integrated intensity between 4.10$^{-3}$ and 3.10$^{-2}$ Å$^{-1}$ (Figure S9) are calculated at different pixel positions in the pouch cell and plotted together with the gas signal and the voltage profile on Figure 2. Graphite interlayer distance increases from the beginning of charge with a strong heterogeneity depending on the position in the pouch cell. Indeed, graphite not directly facing LNO (green) lithiates to a lesser extent than the edges (purple) and center (orange), which is expected due to the oversized Gr-Si electrode compared to LNO. During discharge, almost full graphite delithiation is achieved at the middle of discharge suggesting that only Si delithiation occurs in the second half of discharge. During the 1$^{st}$ charge, the intensity of the crystalline Si peak decreases continuously by up to 35 % (Figure S5 and S6), which is consistent with the crystalline to amorphous transition during Si lithiation and shows the continuous participation of Si during charge. Peaks attributed to Li$_{15}$Si$_4$, the crystalline end-member of the electrochemical Li-Si system, are not evidenced, supporting the incomplete reaction of Si and its participation within an amorphous phase. No changes in the Si peak intensity is observed during discharge showing that the reacted silicon remains amorphous after delithiation. This is a characteristic of Gr-Si composite electrode, for which there is a simultaneous lithiation of graphite and silicon during the first lithiation, and a sequential delithiation with graphite being the first to delithiate[13]. For the LNO, the increase and decrease of *c* parameter during charge appears much more homogenous over the pouch cell with the exception of the H2 → H3 transition, corresponding to the rapid decrease of *c*. Indeed, while the edge is the first to undergo complete H2 → H3 transition, LNO at the reference electrode position still contains H2 after reaching 4.2 V. Delithiation occurs symmetrically to the lithiation. Regarding SAXS data, as the intensity scales as the volume of scatterers, it is very sensitive to local heterogeneities in mass loadings and positions probed in the cell. Therefore, Figure 2 shows the SAXS integrated intensity subtracted from the intensity at the OCV, called normalized SAXS integrated intensity. It remains constant during the formation cycle apart from a reversible increase at the end of charge for the edge and center position. The good correlation between *c* parameter change of LNO and SAXS integrated intensity increase suggests that the SAXS signal is dominated by LNO electrode evolution, potentially linked to the rapid increase of LNO density due to the decrease of the unit cell volume. Gas evolution during formation cycle features $H_2$ and $C_2H_4$ which are observed at the beginning of the first charge along with contributions from $CO_2$ and $C_3H_6$ (Figure S7 and S8) corresponding to the SEI formation[30]. A pronounced $CO_2$ formation starting with the cell charge in Figure S8(d) corresponds to the SEI layer formation and to surface $Li_2CO_3$ oxidation for LNO, in agreement with literature data[31,32]. A small $CO_2$ signal is observed in the lab experiment in Figure S8(d) (not visible during the synchrotron experiment) at the very end of the first charge and at the beginning of discharge likely due to reactive oxygen release from LNO material and reaction with electrolyte as reported previously[14].

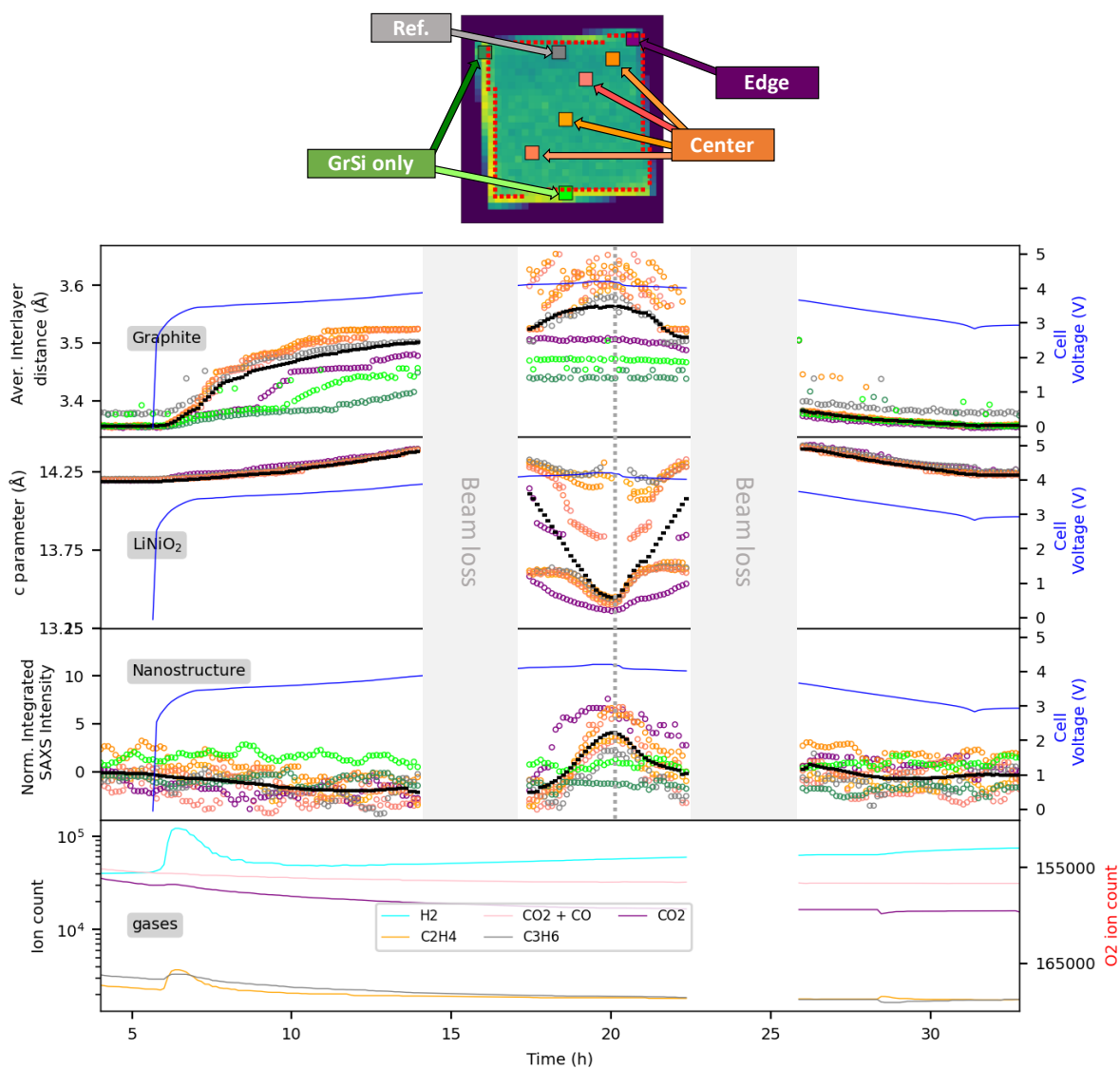

Figure 2: **Structural changes and gas formation during the formation cycle**. From top to bottom, evolution during formation cycle of the interlayer distance in graphite, *c* parameter of $Li_xNiO_2$, SAXS integrated intensity, gas evolution for specific fragments. For all these graphs (apart for the gas analysis), different positions in the pouch cell are shown namely center, reference electrode, edge, and Gr-Si only in orange, brown, purple and green respectively. Black curve corresponds to the average over the entire electrode while the blue curve represents the full cell voltage profile. At the top of the graph, a 2D transmission map of the pouch cell showing selected pixel positions in the pouch cell together with the borders of the LNO electrode (red dotted line). Vertical grey dotted line is a guide to the eye marking the end of the charge.

**Overcharge**

Turning to the overcharge, similar metrics have been extracted from the data and are shown in Figure 3. The general structural evolution during overcharge is similar to the formation cycle with, however a few differences, described and discussed in the following. Focusing first on graphite, lithiation does not start at the beginning of charge but after 2.5 h (at C/9). This results from the lithiation of amorphous silicon formed at the end of the 1st discharge which occurs at higher voltage compared to graphite[33]. At 5 V, there is strong heterogeneities in the interlayer distance across the pouch cell (see further for Li quantification). During the 5 V hold (3h), the heterogeneity does not evolve while the average interlayer distance decreases very slightly suggesting partial delithiation. One of the possible explanations for this can be reactions of intercalated lithium with $CO_2$, as also observed for LFP reference electrode[34,35]. Another possible explanation is a reaction between lithiated graphite and remaining crystalline Si due to charge dynamics redistribution as evidenced by Berhaut *et al.* during relaxation step[36]. Unfortunately, this is difficult to quantify due to the low Si peak intensity. During discharge, and despite the hold, graphite interlayer distance reversibly gets back to initial values, and as observed during the formation cycle, graphite almost fully delithiates during the first half of discharge. For LNO, at 4.3 V, the average *c* parameter reaches 13.34 Å with little in plane heterogeneity indicative of complete H2 → H3 transition. During 5 V hold, the average *c* parameter for LNO decreases slightly showing signs of bulk modification during the hold but without traces of O1 phase formation (Figure S11). These modifications could be related to delithiation of the last Li ions having slow diffusion[37]. During discharge, the average *c* parameter returns back to its original value and for some central position, the H3 phase is observed over a wider voltage range compared to the charge (this will be discussed in more details in next sections).

The overcharge gas evolution is shown Figure 3d, Figure S7-8, which is not described for LNO/Gr-Si cells in the literature yet. There is a massive gas evolution starting at the end of the H2 - H3 transition concomitant with rapid voltage raise. $CO_2$, CO and $O_2$ are produced by the positive electrode side at high voltage. This finding has been confirmed by a specially designed OEMS experiment where a pouch cell containing LNO positive and delithiated $LiFePO_4$ negative electrodes were used, see Figure S10. $CO_2$, CO and $O_2$ are probably due to chemical oxidation of the electrolyte via lattice oxygen as described earlier, and by the direct oxidation of ethylene carbonate, FEC and carbon black at voltages > 4.7 V vs. $Li/Li^+$ [38,4,39]. Note that traces of HF were detected during a 5V hold at m/z = 19, which supports the mechanism of electrochemical oxidation of EC and its subsequent reaction with $PF_6^-$ anion[40]. The other gases observed i.e. $H_2$, $CH_4$, $C_2H_4$ and $C_3H_6$ originate from the reductive decomposition of the electrolyte at the negative electrode at low potentials. The presence of these gases is unexpected as the negative electrode is assumed to be passivated by a SEI layer. Presence of Li plating is ruled out by the high negative/positive electrode loading (N/P) ratio of 1.51 used in this study and the absence of Li metal peaks in the diffraction patterns. The formation of hydrocarbons is therefore indicative of constant SEI formation and breakdown which might originate from the presence of HF and/or gases formed at the cathode damaging the SEI. Note that the amount of formed hydrocarbons is higher than during the SEI formation hence the observed gas evolution is likely not due to residual lithiation of amorphous Si. *Post-mortem* GC-MS analysis of the electrolyte for the cells tested at the synchrotron and in the laboratory revealed considerable consumption of FEC for both cells and *in situ* formation of vinylene carbonate (VC) after cycling, Figure S12 and Table S1. Conversion of FEC to VC was described previously by Etacheri *et al.* which was associated with HF formation [41]. After approx. 1h of 5 V hold, there is an increase of $C_2H_4$ production together with a decrease of the $O_2$, $H_2$ and $CO_2$ gas production which remains constant for the last 2h of the hold. This decrease could be related to a passivation of the LNO oxide surface. This result agrees partially with the data reported by Papp *et al.*[42] in which $CO_2$ evolution was quickly diminished during a 5 V hold for LNO. The concomitant increase of $C_2H_4$ and decrease of $O_2$ can be explained to the exothermic reaction of $C_2H_4$ with $O_2$ forming $CO_2$ [34]. Some of the above-mentioned hypotheses are confirmed by the difference in total amounts of gases generated by LNO/delithiated LFP and LNO/Gr-Si cells during the overcharge cycles. Figure S10(d) demonstrates larger amounts of $CO_2$ and CO produced when Gr-Si electrode was used as well as the absence of $H_2$ and other hydrocarbons for LFP negative electrode. Increased $CO_2$ production for LNO/Gr-Si cell may come from FEC reduction at low potential and from partial conversion of $C_2H_4$ to $CO_2$ in reactions with oxygen[34].

SAXS intensity increases during overcharge in three different regions. First, between 4 V and 4.2 V as observed during the formation cycle, the SAXS intensity increases concomitantly with the H2 → H3 transition. From 4.2 V to 5 V, the unit cell volume of LNO barely changes, while the SAXS intensity continues to increase. Interestingly, this voltage range is exactly where gas evolution starts to increase massively. During the hold, the SAXS intensity continues to increase with a slower rate while the gas evolution rate also slows down. At the end of discharge, the intensity does not come back to its original value showing the irreversibility of some of the probed nanostructural changes. Analysing quantitatively the SAXS signal is difficult, as it contains contributions from both anodic and cathodic materials, arising from large grains interfaces as well as nanosized objects. Morevoer, it scales with different terms as volume fraction, specific surface, two-phase electronic contrast and particle form factors. Nevertheless, the correlation between the observed LNO structural changes is a strong indication that the the SAXS intensity increase could be due to densification of the LNO due to 1) massive *c* parameter contraction during the H2 → H3 transition, and 2) the irreversible formation of densified surface layer due to O loss (See SI for more details).

Before moving on to next section, we summarize some of the important aspects of LNO/Gr-Si cell overcharge revealed by the direct correlation between structural evolution and gas released: (1) massive gas release from the LNO can be observed without substantial O1 phase formation, (2) gas released can react with cell components (oxidizing the electrolyte, LFP, or lithiated graphite as observed by the reduction of the interlayer distance during hold), (3) overcharge also leads to substantial gas release from the anode even if it is not fully lithiated and in absence of Li plating, (4) there is concomitant decrease of the nanostructural change rate and gas evolution after 1 h of hold at 5 V which is probably due to a close to complete surface densification on LNO.

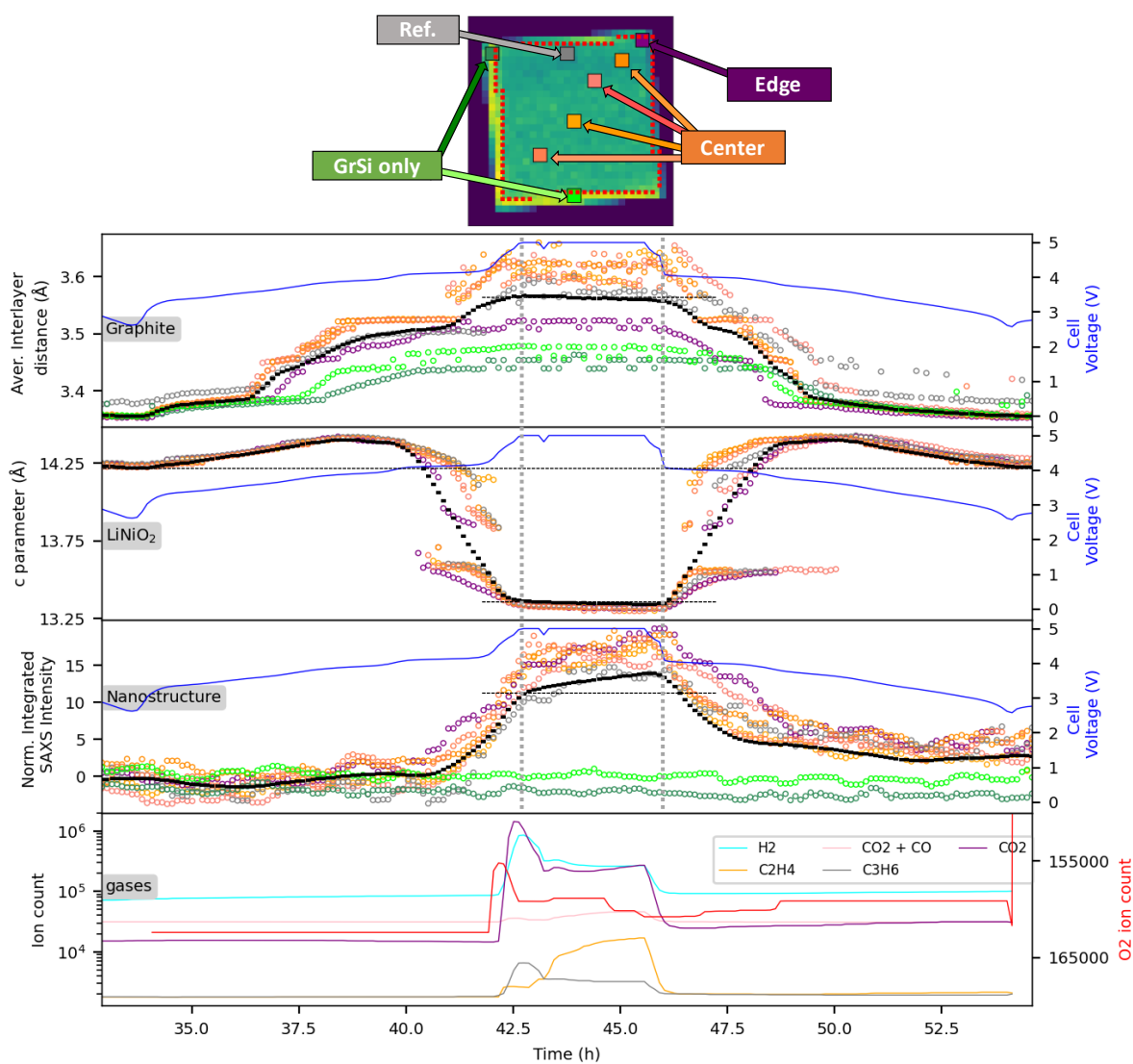

Figure 3: **Structural changes and gas formation during the overcharge cycle**. Evolution during overcharge of the interlayer distance in graphite, *c* parameter of Li$_x$NiO$_2$, SAXS integrated intensity, gas evolution for specific fragments. For all these graphs (apart for the gas analysis), different positions in the pouch cell are shown namely center, ref, edge, and GrSi only in orange, brown, purple and green respectively. Black curve corresponds to the average over the entire electrode while the blue curve represents the full cell voltage profile. At the top of the graph, a transmission image of the pouch cell showing the selected positions in the pouch cell together with the borders of the LNO electrode (red dotted line). Horizontal lines at guide to the eyes to better appreciate to lattice parameter and SAXS intensity change during the hold at 5 V. Vertical grey dotted lines indicate the end of the charge (before the hold) and the beginning of the discharge.

Resolving spatial heterogeneities

From the previous section, it appears that the average structural evolution of LNO and graphite during the discharge after overcharge is fairly similar to the discharge of the formation cycle, hence suggesting only minor impacts of overcharge on the battery performance. To get more details about the effect of the overcharge, we analyze the spatial distribution of lithiation rates during and after the overcharge. Along that line, Li concentration maps in both electrode materials for all times have been produced based on lattice parameter variations. Briefly, $a_{LNO}$ parameter is determined from both (101) and (003) reflections and used to estimate lithium concentration in LNO using charge of the formation cycle and reference data from literature (Figure S13 and S14)[14]. On the graphite side, lithium concentration is estimated from both the phase fraction and interlayer distance of the different stages as reported earlier[27]. After obtaining dynamic lithium concentration maps, heterogeneity maps are produced by subtracting the spatial average across the entire electrode from the local Li concentration ([Li] in a pixel) (Figure 4 and 5 for LNO and Gr respectively). Note that for graphite, the average Li concentration was not calculated over the entire electrode but over the electrode part directly facing LNO electrode. This allows better visualization of heterogeneities in the electrochemical stack by ruling out the effect of the oversized anode electrode. In the following, we start by describing the heterogeneity of the lithiation rate followed by a discussion on the origin of such phenomena.

Focusing on the overcharge of LNO, the heterogeneity profile (spatial distribution of heterogeneity) depends on the state of charge and is visually maximum at approximately x = 0.75, as observed on the map at x = 0.76 shown Figure 4a. For x ranging between 1 and 0.5 (x in $Li_xNiO_2$), electrode edges are always more delithiated (red) compared to the center of the pouch cell (blue). At higher SoC (0 < x < 0.5), the heterogeneities are weaker and the bottom edge of the pouch cell is the most delithiated zone. This edge was held down during the experiment and is also where the positive and negative tabs are located. It suggests that either electrolyte accumulation and/or shorter electronic diffusion pathways could results from lower overpotential hence early delithiation of LNO. At high state of charge, heterogeneities can be observed at the reference electrode position (blue dots at x = 0.32). During discharge after the overcharge, the general heterogeneity profile is similar to the charge (LNO in center of pouch cell is more lithiated compared to the edge) with however the presence of some very local heterogeneities not observed during overcharge and located at the center of the pouch cell, or close to the reference electrode position for example (visualized on Figure 4b as red pixels hence more delithiated areas). To quantitatively measure the intensity of the heterogeneities, standard deviation of the Li concentration histogram in the LNO electrode is shown in Figure 4c for all cycles. Apart from the 1st charge (See Figure S15 and S16), the heterogeneity versus SoC curve shows several minima located approx. at x = 0.85, 0.65, 0.5, 0.3. Interestingly, there is a correlation between these minima and dQ/dV of LNO (Figure 4d) indicating that heterogeneities arise at voltage plateaus. Discharges have similar profiles and are more heterogeneous than the overcharge (i.e. shows a higher standard deviation value).

Turning to the graphite during overcharge (Figure 5), the heterogeneity profile clearly shows that graphite not facing LNO electrode display a lower degree of lithiation (red). In the following, we focus on the heterogeneity of graphite facing LNO electrode (that is inside the black dotted line representing the LNO electrode contour in Figure 5). Generally, graphite is more lithiated (blue) in the center of the pouch cell with the exception of a broad horizontal line crossing the pouch cell. Lithiation difference between edges and center is SoC dependent, as it is stronger for example at x = 0.39 compared to x = 0.49. During discharge, the same heterogeneity pattern is observed with also very local heterogeneities appearing as speckles on map at x = 0.30 or blue dots at x = 0.09 located at the center of the pouch and reference electrode (more lithiated areas). On the standard deviation of the Li concentration histogram for the graphite directly facing LNO presented in Figure 5, it can be seen that (1) heterogeneity values are in average twice as high in graphite compared to LNO, (2) the heterogeneity depends on SoC with local minima at x ≈ 0.1, and 0.5 somewhat matching the inflexion points in graphite electrochemical curve during lithiation (corresponding to pure Stage3L and Stage2), (3) the discharge after the overcharge is much more heterogeneous especially for 0 < x < 0.4.

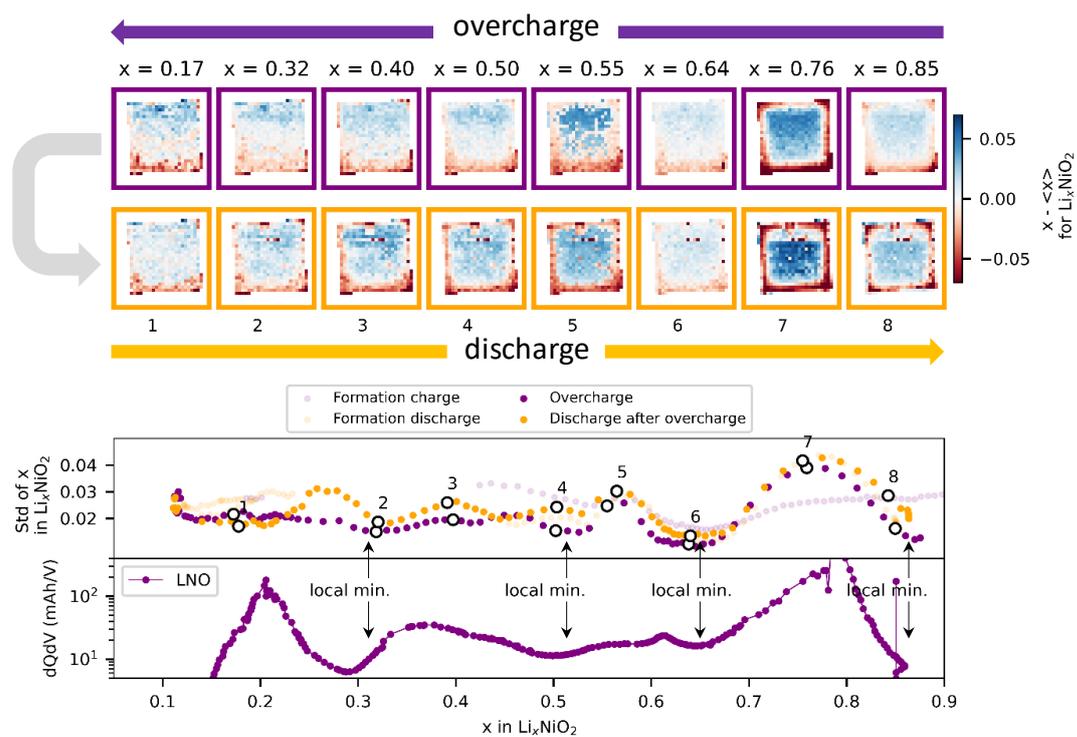

Figure 4: **Reaction heterogeneity in LNO during overcharge and the discharge after overcharge**. Top panel shows Li concentration deviation maps for the overcharge (top) and discharge after overcharge (bottom). Li concentration deviation maps are calculated by subtracting the spatial average across the entire electrode, <x>, from the local Li concentration, x, resulting in blue and red regions corresponding to zones having more and less Li compared to the average electrode. Maps are shown at key steps labeled 1 to 8, ranging from <x>=0.85 to <x>=0.17. The bottom panel shows the standard deviation of the Li concentration histogram for each maps – which is a quantitative description of the amplitude of the spatial heterogeneity – together with the dQ/dVdV of the overcharge cycle calculated for the LNO. Purple colors are used for the charges and orange for the discharges. Note that the charge starts at $Li_{0.85}NiO_2$ due to Li loss in SEI formation during the formation cycle

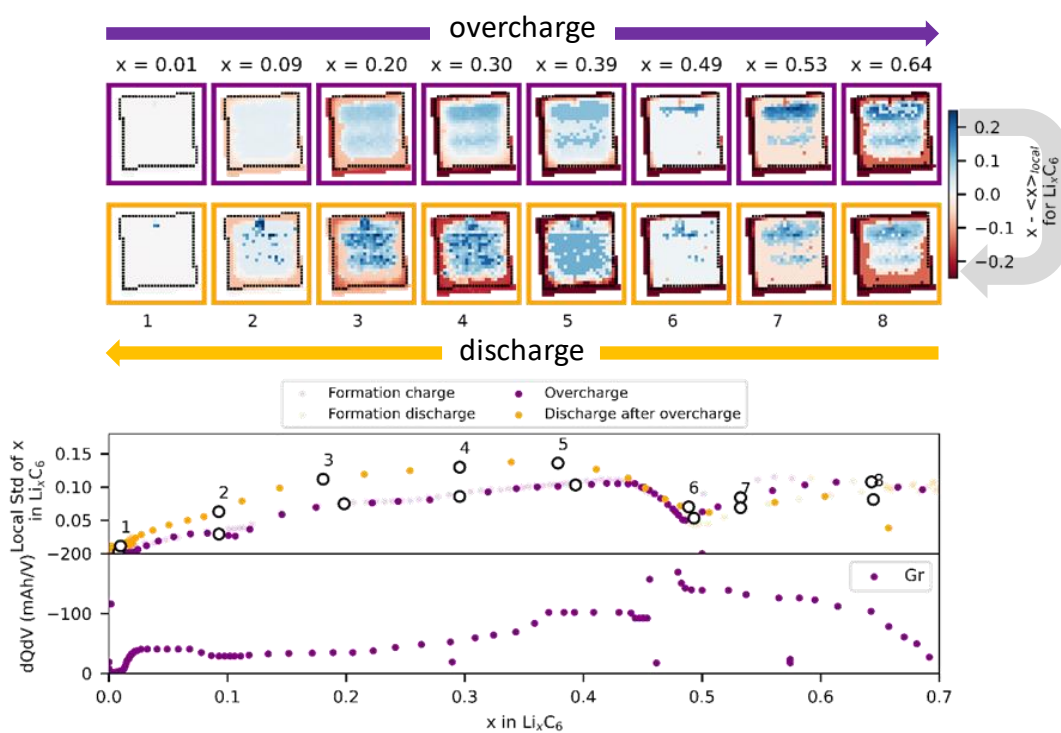

Figure 5: **Reaction heterogeneity in graphite during overcharge and the discharge after overcharge**. Top panel shows Li concentration deviation maps for the overcharge (top) and discharge after overcharge (bottom). Li concentration deviation maps are calculated as follow : for each pixel, the local Li concentration (of the pixel), x, is subtracted the average Li concentration in the graphite electrode facing the LNO electrode, $<x>_{local}$, resulting in red and blue regions corresponding to zones having more and less Li compared to the average electrode. The bottom panel shows the standard deviation of the Li concentration histogram for each maps – which is quantitative description of the amplitude of the spatial heterogeneity – together with the dQ/dVdV of the overcharge cycle calculated for the graphite-silicon electrodes. Purple colors are used for the charges and orange for the discharges.

We have shown that overcharge leads to local heterogeneities during the subsequent discharge. Interestingly, some of these local heterogeneities are found at the same positions on both positive and negative electrodes, as highlighted in Figure 6a-b. Moreover, by comparing Li concentration in LNO and graphite at these positions with electrode ensemble Li concentration (Figure 6c), it is clear that reactions at these positions are kinetically limited. Indeed, for LNO the local Li concentration during overcharge (purple line) matches the ensemble electrode lithiation as it follows the straight dashed lines (corresponding to $x_{local} = x_{ensemble}$). However, during the subsequent discharge, the local Li concentration (orange line) during the discharge deviate from the dashed line with the local concentration being smaller compared to the ensemble electrode concentration. For the graphite, local Li concentration during overcharging does not superimpose to the dashed line (due to the much larger heterogeneity observed in the graphite electrode compared to LNO), but the deviation to the ensemble Li concentration is even more pronounced during discharge. Note that before 5 V hold, (de)lithiation kinetics at these positions was closer to the ensemble electrode reaction confirming the detrimental effect of overcharge. Interestingly, the pouch cell was opened after the experiment and a visible defect was observed in the anode, which position matches some of the most pronounced damaged zones (Figure 6d).

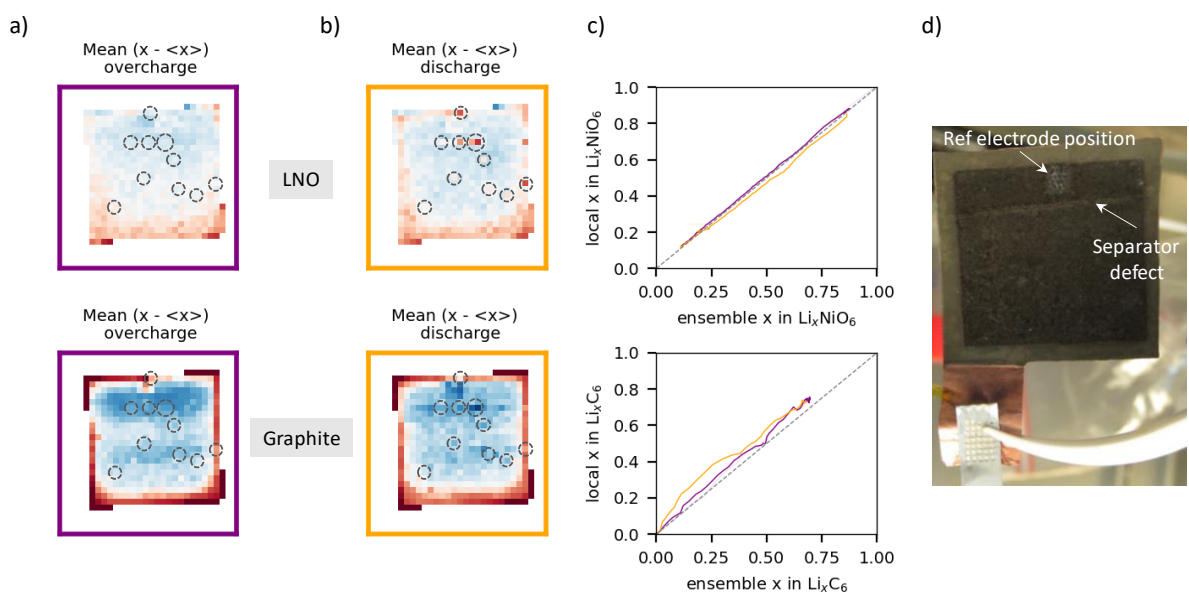

Figure 6: **Overcharge based local heterogeneities**. a) and b) are Li concentration deviation maps averaged over the entire overcharge and discharge, respectively (shown in purple and orange, respectively). Top and bottom maps are for LNO and graphite, respectively. All maps have dotted circles highlighting the position of heterogeneities observed during the overcharge. Dotted circles are the same for all conditions and sample. c) Electrode average Li concentration (ensemble x in $Li_xNiO_2$ and $Li_xC_6$ for the top and bottom panel, respectively) versus local Li concentration inside the circles shown in the maps a) and b). Purple and orange lines are for the overcharge and discharge, respectively. d) Picture of the dried and washed negative electrode after disassembling of the pouch with traces from the reference electrode and from the separator defect (line also visible on the separator after disassembling the cell).

Discussion on the origin of the heterogeneities:

In summary, we evidence four heterogeneity effects: (1) graphite not facing LNO is under lithiated, (2) there is frequently a delay of the reaction mechanism close to the reference which is presumably due to mechanical deformation leading to different current distribution (3) lithiation heterogeneity between electrodes edge and center is SoC dependent and occurs at every cycle, (4) there are local spots lagging behind the ensemble electrochemistry only present during the discharge after the overcharge.

In the following, we discuss the origins of the third and four type of heterogeneity which have very different characteristics. Regarding the SoC-dependent heterogeneity (n°3), the edges (4-5 mm) of LNO are always more delithiated compared to the center of the pouch cell in charge or discharge, or during voltage holds. For graphite, the edges are also more delithiated during charge and discharge together with a broad line visible at the center of the pouch cell. These observations shows that Li inventory (that is the sum of Li in cathode and anode) might not be spatially homogeneous over mm (in this case, area larger than the mm at the edges have less lithium compared to the center). Moreover, our observation also shows that this type of in plane heterogeneity is not kinetically controlled but rather related to the shape of the voltage curve. To explain why both the LNO and graphite electrode are under lithiated at the edges, we propose that Li extracted from LNO inserts over a large region in the graphite electrode (extending to graphite not directly facing LNO electrode), which leads to under lithiated graphite. In turn, the under lithiated graphite electrochemical potential is higher which drives a higher electrochemical potential at the LNO electrode because full cell voltage is fixed. The raise of the electrochemical potential at the edges leads to lower Li concentration in LNO especially around flat voltage regions for which a small change in potential leads to large Li concentration change. This explains the SoC dependence of the heterogeneity. This effect, emphasized in our single layer pouch cell, may also appear in commercial system due to other sources of local heterogeneous cell balancing. The fourth heterogeneity is only visible after the overcharge, and consists of very local spots which appear to be lagging behind the ensemble electrochemical reaction (in this case, slower delithiation for LNO and lithiation for graphite). Some of these local spots are visible for both electrodes while some are only observed on the graphite electrode – which seems more affected by overcharge based on the standard deviation values reported in Figures 4 and 5. For the spots present on both electrodes and matching the presence of visible separator defects, we believe the delay in the electrochemical reaction might be caused by gas bubbles gathering at these positions, coalescing and pushing out the electrolyte, hence leading to higher resistance areas. Interestingly the rest of the LNO electrode does not seem to be affected by overcharge despite massive gas release and nanostructural modification at relatively slow C-rate used in this work. Graphite is more affected which is possibly correlated with the presence of $C_2H_4$ and $H_2$ suggesting SEI evolution on graphite surface.

**Conclusion :** In this work, we have demonstrated the proof of concept of an *operando* multi-probe experiment producing high-quality and correlated WAXS/SAXS mapping together with gas evolution data during formation and overcharge of $LiNiO_2$/Gr-Si single layer pouch cell without beam damage effect on any of the datasets. Correlative analysis of the datasets provide a series of new insights on overcharge mechanism namely (1) massive gas release from the LNO can be observed without substantial O1 phase formation, (2) gas released can react with cell components (oxidizing the LFP or graphite-silicon electrode) questioning the general perspective on the chemical durability of sensors in battery, (3) overcharge leads to substantial gas release from the anode even if it is not fully lithiated and in absence of Li plating showing the importance of cross talks reactions, (4) defects, tabs, ref electrode and wrong local electrode balancing introduces heterogeneous reactions which needs to be taken into consideration for thermal and/or ageing simulations (4) overcharge produces local spots featuring kinetically limited (de)lithiation reaction possibly due to trapping of gas bubbles in cell manufacturing defects. This method can be applied to investigate more severe overcharging conditions and/or high temperatures, as requested for battery homologation. It would be particularly interesting to follow Ni-rich materials thermal stability in a variety of usage or abusive conditions, as it is directly linked to crystal changes and oxygen release. Moreover, our results emphasize the importance of controlling internal defects that may induce thermal, mechanical and electrical heterogeneities affecting local chemistry and electrochemical redox reactions. Homogeneous cell designs optimizing tabs location, or smart cells built with miniaturized gas sensors and stable reference electrode, together with chemistry tuning (Ni-rich coating, electrolyte,…), appear to be key to limit gassing and avoid the local increase of internal resistance, inducing local overheating and over(de)lithiation, and potentially leading to safety issues. Coupling the OEMS-SAXS/WAXS set-up to other probes, particularly to chemical-sensitive techniques capable to measure soluble species or interfacial compounds, would be the next step to integrate knowledge from electrochemistry, structure, gas generation and chemical environment measures for a full holistic characterization of full cells.

**Acknowledgements:** Beamtime at the ESRF was granted within the Battery Pilot Hub MA4929 "Multi-scale Multi-techniques investigations of Li-ion batteries: towards a European Battery Hub". We acknowledge support from EU H2020 project BIG-MAP (grant agreement ID: 957189). We are grateful to J.F. Martin for conducting post-mortem GC-MS analysis of the electrolytes and to D. Buzon for fruitful discussions.

**Author contributions :** Q.J. and I.P. conceived the idea. S.L. supervised the project. Q.J., S.T. and S.L. designed the synchrotron experiment. I.P., L.B., E.A., B.A., P.C. prepared the samples and performed the lab scale experiments. N.B. aligned and set the beamline. All authors participated to the beamtime except of L.D. I.P. and L.B. analyzed the gas data, Q.J. and S.L. analyzed X-ray data, Q.J, I.P. L.B. wrote the manuscript, L.D. and S.L. revised the manuscript, all authors discussed the results and contributed to the manuscript.

**Competing interests:** The authors declare no competing interests.

**References :**
(1) Vidal, O.; Goffé, B.; Arndt, N. Metals for a Low-Carbon Society. *Nature Geosci* **2013**, *6* (11), 894–896. https://doi.org/10.1038/ngeo1993.
(2) Herrington, R. Mining Our Green Future. *Nat Rev Mater* **2021**, *6* (6), 456–458. https://doi.org/10.1038/s41578-021-00325-9.
(3) Bianchini, M.; Roca-Ayats, M.; Hartmann, P.; Brezesinski, T.; Janek, J. There and Back Again—The Journey of LiNiO2 as a Cathode Active Material. *Angewandte Chemie International Edition* **2019**, *58* (31), 10434–10458. https://doi.org/10.1002/anie.201812472.
(4) Cui, Z.; Manthiram, A. Thermal Stability and Outgassing Behaviors of High-Nickel Cathodes in Lithium-Ion Batteries. *Angewandte Chemie International Edition* **2023**, *62* (43), e202307243. https://doi.org/10.1002/anie.202307243.
(5) Pan, R.; Jo, E.; Cui, Z.; Manthiram, A. Degradation Pathways of Cobalt-Free LiNiO2 Cathode in Lithium Batteries. *Advanced Functional Materials* **2023**, *33* (10), 2211461. https://doi.org/10.1002/adfm.202211461.
(6) Kalaga, K.; Rodrigues, M.-T. F.; Trask, S. E.; Shkrob, I. A.; Abraham, D. P. Calendar-Life versus Cycle-Life Aging of Lithium-Ion Cells with Silicon-Graphite Composite Electrodes. *Electrochimica Acta* **2018**, *280*, 221–228. https://doi.org/10.1016/j.electacta.2018.05.101.
(7) Metzger, M.; Strehle, B.; Solchenbach, S.; Gasteiger, H. A. Origin of H2 Evolution in LIBs: H2O Reduction vs. Electrolyte Oxidation. *J. Electrochem. Soc.* **2016**, *163* (5), A798. https://doi.org/10.1149/2.1151605jes.
(8) Müller, S.; Eller, J.; Ebner, M.; Burns, C.; Dahn, J.; Wood, V. Quantifying Inhomogeneity of Lithium Ion Battery Electrodes and Its Influence on Electrochemical Performance. *J. Electrochem. Soc.* **2018**, *165* (2), A339. https://doi.org/10.1149/2.0311802jes.
(9) Zhang, L.; Huang, L.; Zhang, Z.; Wang, Z.; Dorrell, D. D. Degradation Characteristics Investigation for Lithium-Ion Cells with NCA Cathode during Overcharging. *Applied Energy* **2022**, *327*, 120026. https://doi.org/10.1016/j.apenergy.2022.120026.
(10) S. Edge, J.; O'Kane, S.; Prosser, R.; D. Kirkaldy, N.; N. Patel, A.; Hales, A.; Ghosh, A.; Ai, W.; Chen, J.; Yang, J.; Li, S.; Pang, M.-C.; Diaz, L. B.; Tomaszewska, A.; Waseem Marzook, M.; N. Radhakrishnan, K.; Wang, H.; Patel, Y.; Wu, B.; J. Offer, G. Lithium Ion Battery Degradation: What You Need to Know. *Physical Chemistry Chemical Physics* **2021**, *23* (14), 8200–8221. https://doi.org/10.1039/D1CP00359C.
(11) Atkins, D.; Capria, E.; Edström, K.; Famprikis, T.; Grimaud, A.; Jacquet, Q.; Johnson, M.; Matic, A.; Norby, P.; Reichert, H.; Rueff, J.-P.; Villevieille, C.; Wagemaker, M.; Lyonnard, S. Accelerating Battery Characterization Using Neutron and Synchrotron Techniques: Toward a Multi-Modal and Multi-Scale Standardized Experimental Workflow. *Advanced Energy Materials* **2022**, *12* (17), 2102694. https://doi.org/10.1002/aenm.202102694.

(12) Dawkins, J. I. G.; Martens, I.; Danis, A.; Beaulieu, I.; Chhin, D.; Mirolo, M.; Drnec, J.; Schougaard, S. B.; Mauzeroll, J. Mapping the Total Lithium Inventory of Li-Ion Batteries. *Joule* **2023**, *7* (12), 2783–2797. https://doi.org/10.1016/j.joule.2023.11.003.

(13) Berhaut, C. L.; Dominguez, D. Z.; Kumar, P.; Jouneau, P.-H.; Porcher, W.; Aradilla, D.; Tardif, S.; Pouget, S.; Lyonnard, S. Multiscale Multiphase Lithiation and Delithiation Mechanisms in a Composite Electrode Unraveled by Simultaneous Operando Small-Angle and Wide-Angle X-Ray Scattering. *ACS Nano* **2019**, *13* (10), 11538–11551. https://doi.org/10.1021/acsnano.9b05055.

(14) de Biasi, L.; Schiele, A.; Roca-Ayats, M.; Garcia, G.; Brezesinski, T.; Hartmann, P.; Janek, J. Phase Transformation Behavior and Stability of LiNiO2 Cathode Material for Li-Ion Batteries Obtained from In Situ Gas Analysis and Operando X-Ray Diffraction. *ChemSusChem* **2019**, *12* (10), 2240–2250. https://doi.org/10.1002/cssc.201900032.

(15) Sim, R.; Langdon, J.; Manthiram, A. Design of an Online Electrochemical Mass Spectrometry System to Study Gas Evolution from Cells with Lean and Volatile Electrolytes. *Small Methods* **2023**, *7* (6), 2201438. https://doi.org/10.1002/smtd.202201438.

(16) Juelsholt, M.; Chen, J.; A. Pérez-Osorio, M.; J. Rees, G.; Coutinho, S. D. S.; E. Maynard-Casely, H.; Liu, J.; Everett, M.; Agrestini, S.; Garcia-Fernandez, M.; Zhou, K.-J.; A. House, R.; G. Bruce, P. Does Trapped O$_2$ Form in the Bulk of LiNiO$_2$ during Charging? *Energy & Environmental Science* **2024**. https://doi.org/10.1039/D3EE04354A.

(17) Chien, P.-H.; Wu, X.; Song, B.; Yang, Z.; Waters, C. K.; Everett, M. S.; Lin, F.; Du, Z.; Liu, J. New Insights into Structural Evolution of LiNiO2 Revealed by Operando Neutron Diffraction. *Batteries & Supercaps* **2021**, *4* (11), 1701–1707. https://doi.org/10.1002/batt.202100135.

(18) Xu, C.; Reeves, P. J.; Jacquet, Q.; Grey, C. P. Phase Behavior during Electrochemical Cycling of Ni-Rich Cathode Materials for Li-Ion Batteries. *Advanced Energy Materials* **2021**, *11* (7), 2003404. https://doi.org/10.1002/aenm.202003404.

(19) Croguennec, L.; Pouillerie, C.; Mansour, A. N.; Delmas, C. Structural Characterisation of the Highly Deintercalated LixNi1.02O2 Phases (with x ≤ 0.30). *J. Mater. Chem.* **2001**, *11* (1), 131–141. https://doi.org/10.1039/b003377o.

(20) Park, K.; Zhu, Y.; Torres-Castanedo, C. G.; Jung, H. J.; Luu, N. S.; Kahveci̇oglu, O.; Yoo, Y.; Seo, J. T.; Downing, J. R.; Lim, H.; Bedzyk, M. J.; Wolverton, C.; Hersam, M. C. Elucidating and Mitigating High-Voltage Degradation Cascades in Cobalt-Free LiNiO$_2$ Lithium-Ion Battery Cathodes. *Advanced Materials* **2022**, *34* (3), 2106402. https://doi.org/10.1002/adma.202106402.

(21) P. Paul, P.; Thampy, V.; Cao, C.; Steinrück, H.-G.; R. Tanim, T.; R. Dunlop, A.; J. Dufek, E.; E. Trask, S.; N. Jansen, A.; F. Toney, M.; Weker, J. N. Quantification of Heterogeneous, Irreversible Lithium Plating in Extreme Fast Charging of Lithium-Ion Batteries. *Energy & Environmental Science* **2021**, *14* (9), 4979–4988. https://doi.org/10.1039/D1EE01216A.

(22) Graae, K. V.; Li, X.; Sørensen, D. R.; Ayerbe, E.; Boyano, I.; Sheptyakov, D.; Jørgensen, M. R. V.; Norby, P. Time and Space Resolved Operando Synchrotron X-Ray and Neutron Diffraction Study of NMC811/Si–Gr 5 Ah Pouch Cells. *Journal of Power Sources* **2023**, *570*, 232993. https://doi.org/10.1016/j.jpowsour.2023.232993.

(23) P. Finegan, D.; Quinn, A.; S. Wragg, D.; M. Colclasure, A.; Lu, X.; Tan, C.; M. Heenan, T. M.; Jervis, R.; L. Brett, D. J.; Das, S.; Gao, T.; A. Cogswell, D.; Z. Bazant, M.; Michiel, M. D.; Checchia, S.; R. Shearing, P.; Smith, K. Spatial Dynamics of Lithiation and Lithium Plating during High-Rate Operation of Graphite Electrodes. *Energy & Environmental Science* **2020**, *13* (8), 2570–2584. https://doi.org/10.1039/D0EE01191F.

(24) Kieffer, J.; Karkoulis, D. PyFAI, a Versatile Library for Azimuthal Regrouping. *J. Phys.: Conf. Ser.* **2013**, *425* (20), 202012. https://doi.org/10.1088/1742-6596/425/20/202012.

(25) Flores, E.; Mozhzhukhina, N.; Li, X.; Norby, P.; Matic, A.; Vegge, T. PRISMA: A Robust and Intuitive Tool for High-Throughput Processing of Chemical Spectra**. *Chemistry–Methods* **2022**, *2* (10), e202100094. https://doi.org/10.1002/cmtd.202100094.

(26) Didier, C.; Pang, W. K.; Guo, Z.; Schmid, S.; Peterson, V. K. Phase Evolution and Intermittent Disorder in Electrochemically Lithiated Graphite Determined Using in Operando Neutron Diffraction. *Chem. Mater.* **2020**, *32* (6), 2518–2531. https://doi.org/10.1021/acs.chemmater.9b05145.

(27) Tardif, S.; Dufour, N.; Colin, J.-F.; Gébel, G.; Burghammer, M.; Johannes, A.; Lyonnard, S.; Chandesris, M. Combining Operando X-Ray Experiments and Modelling to Understand the Heterogeneous Lithiation of Graphite Electrodes. *Journal of Materials Chemistry A* **2021**, *9* (7), 4281–4290. https://doi.org/10.1039/D0TA10735B.

(28) Wetjen, M.; Solchenbach, S.; Pritzl, D.; Hou, J.; Tileli, V.; Gasteiger, H. A. Morphological Changes of Silicon Nanoparticles and the Influence of Cutoff Potentials in Silicon-Graphite Electrodes. *J. Electrochem. Soc.* **2018**, *165* (7), A1503. https://doi.org/10.1149/2.1261807jes.

(29) Jousseaume, T.; Colin, J.-F.; Chandesris, M.; Lyonnard, S.; Tardif, S. Strain and Collapse during Lithiation of Layered Transition Metal Oxides: A Unified Picture. *Energy & Environmental Science* **2024**. https://doi.org/10.1039/D3EE04115H.

(30) Rowden, B.; Garcia-Araez, N. A Review of Gas Evolution in Lithium Ion Batteries. *Energy Reports* **2020**, *6*, 10–18. https://doi.org/10.1016/j.egyr.2020.02.022.

(31) Jung, R.; Metzger, M.; Maglia, F.; Stinner, C.; Gasteiger, H. A. Oxygen Release and Its Effect on the Cycling Stability of LiNixMnyCozO2 (NMC) Cathode Materials for Li-Ion Batteries. *J. Electrochem. Soc.* **2017**, *164* (7), A1361. https://doi.org/10.1149/2.0021707jes.

(32) Kaufman, L. A.; McCloskey, B. D. Surface Lithium Carbonate Influences Electrolyte Degradation via Reactive Oxygen Attack in Lithium-Excess Cathode Materials. *Chem. Mater.* **2021**, *33* (11), 4170–4176. https://doi.org/10.1021/acs.chemmater.1c00935.

(33) Krause, L. J.; Brandt, T.; Chevrier, V. L.; Jensen, L. D. Surface Area Increase of Silicon Alloys in Li-Ion Full Cells Measured by Isothermal Heat Flow Calorimetry. *J. Electrochem. Soc.* **2017**, *164* (9), A2277. https://doi.org/10.1149/2.0501712jes.

(34) Ellis, L. D.; Allen, J. P.; Thompson, L. M.; Harlow, J. E.; Stone, W. J.; Hill, I. G.; Dahn, J. R. Quantifying, Understanding and Evaluating the Effects of Gas Consumption in Lithium-Ion Cells. *J. Electrochem. Soc.* **2017**, *164* (14), A3518. https://doi.org/10.1149/2.0191714jes.

(35) Sloop, S. E.; Kerr, J. B.; Kinoshita, K. The Role of Li-Ion Battery Electrolyte Reactivity in Performance Decline and Self-Discharge. *Journal of Power Sources* **2003**, *119–121*, 330–337. https://doi.org/10.1016/S0378-7753(03)00149-6.

(36) Berhaut, C. L.; Mirolo, M.; Dominguez, D. Z.; Martens, I.; Pouget, S.; Herlin-Boime, N.; Chandesris, M.; Tardif, S.; Drnec, J.; Lyonnard, S. Charge Dynamics Induced by Lithiation Heterogeneity in Silicon-Graphite Composite Anodes. *Advanced Energy Materials* **2023**, *13* (44), 2301874. https://doi.org/10.1002/aenm.202301874.

(37) Märker, K.; Reeves, P. J.; Xu, C.; Griffith, K. J.; Grey, C. P. Evolution of Structure and Lithium Dynamics in LiNi0.8Mn0.1Co0.1O2 (NMC811) Cathodes during Electrochemical Cycling. *Chem. Mater.* **2019**, *31* (7), 2545–2554. https://doi.org/10.1021/acs.chemmater.9b00140.

(38) Rinkel, B. L. D.; Vivek, J. P.; Garcia-Araez, N.; Grey, C. P. Two Electrolyte Decomposition Pathways at Nickel-Rich Cathode Surfaces in Lithium-Ion Batteries. *Energy Environ. Sci.* **2022**, *15* (8), 3416–3438. https://doi.org/10.1039/D1EE04053G.

(39) Xia, L.; Tang, B.; Yao, L.; Wang, K.; Cheris, A.; Pan, Y.; Lee, S.; Xia, Y.; Chen, G. Z.; Liu, Z. Oxidation Decomposition Mechanism of Fluoroethylene Carbonate-Based Electrolytes for High-Voltage Lithium Ion Batteries: A DFT Calculation and Experimental Study. *ChemistrySelect* **2017**, *2* (24), 7353–7361. https://doi.org/10.1002/slct.201700938.

(40) Solchenbach, S.; Metzger, M.; Egawa, M.; Beyer, H.; Gasteiger, H. A. Quantification of PF5 and POF3 from Side Reactions of LiPF6 in Li-Ion Batteries. *J. Electrochem. Soc.* **2018**, *165* (13), A3022. https://doi.org/10.1149/2.0481813jes.

(41) Etacheri, V.; Haik, O.; Goffer, Y.; Roberts, G. A.; Stefan, I. C.; Fasching, R.; Aurbach, D. Effect of Fluoroethylene Carbonate (FEC) on the Performance and Surface Chemistry of Si-Nanowire Li-Ion Battery Anodes. *Langmuir* **2012**, *28* (1), 965–976. https://doi.org/10.1021/la203712s.

(42) Papp, J. K.; Li, N.; Kaufman, L. A.; Naylor, A. J.; Younesi, R.; Tong, W.; McCloskey, B. D. A Comparison of High Voltage Outgassing of LiCoO2, LiNiO2, and Li2MnO3 Layered Li-Ion Cathode Materials. *Electrochimica Acta* **2021**, *368*, 137505. https://doi.org/10.1016/j.electacta.2020.137505.

Supporting information for:

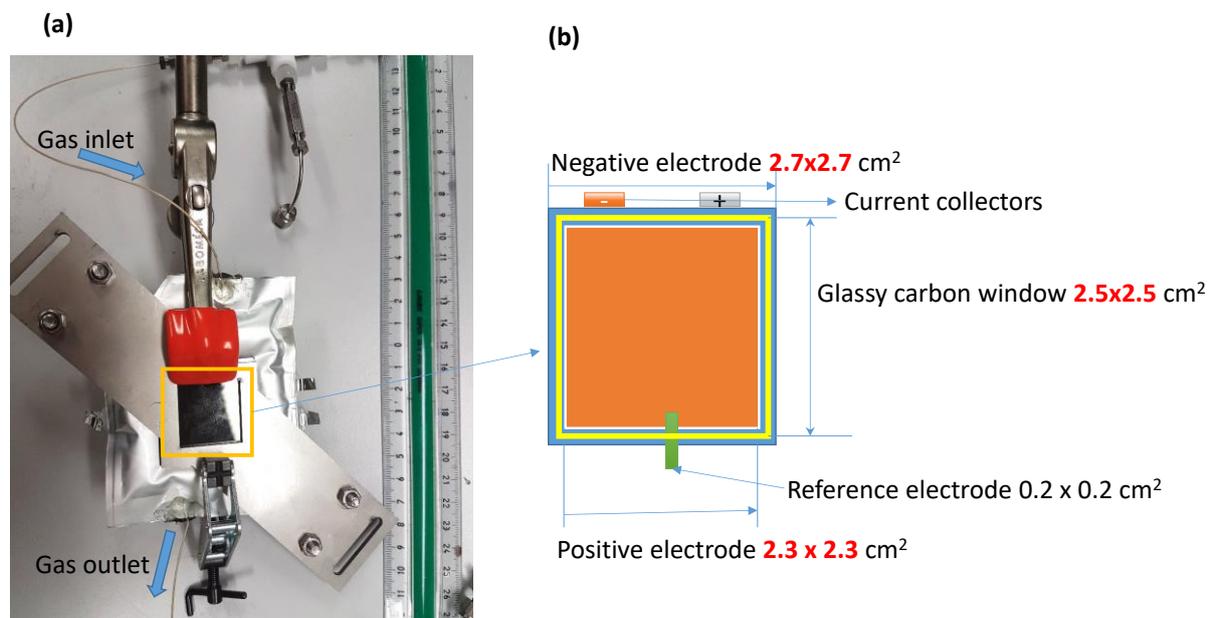

Figure S1: A photograph of the instrumented pouch cell tested during this study with glassy carbon windows and a stainless steel holder/frame (a); schematic view of the electrode stack with dimensions of the electrodes and the glassy carbon window transparent for SAXS/WAXS.

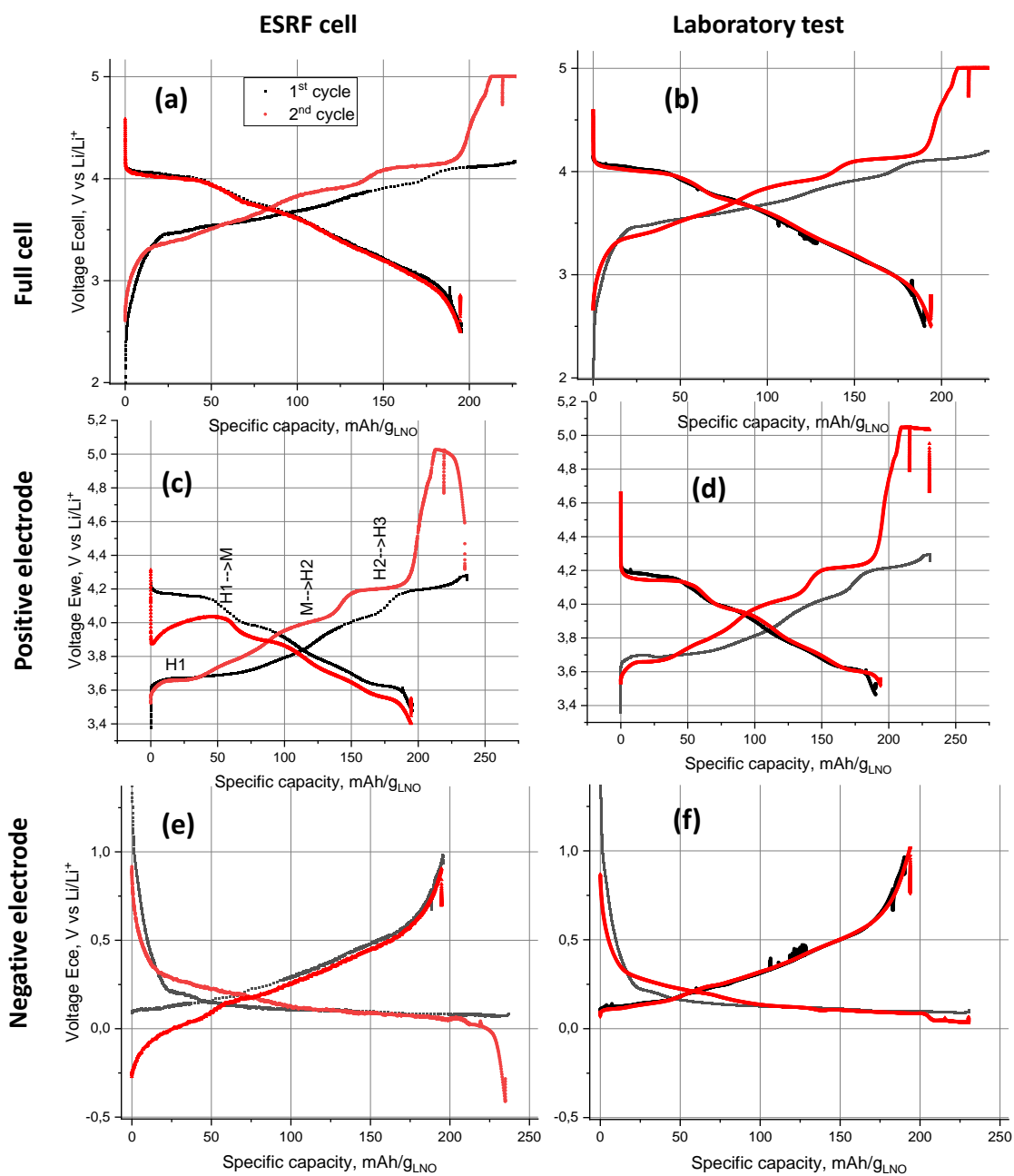

Figure S2. Electrochemical charge –discharge curves obtained during the two first cycles with different upper cut-off limits for two identical LNO//GrSi- cells with partially delithiated LFP reference electrodes in EC/1.3M LiPF$_6$, 10%wt. FEC electrolyte tested at ESRF (a), (c), (e) and outside the beamline in the laboratory (b), (d) and (f). Full cell voltage profiles (a) and (b); positive electrode potential profiles measured with the help of the reference electrode (c) and (d); and negative electrode potential profiles (e) and (f).

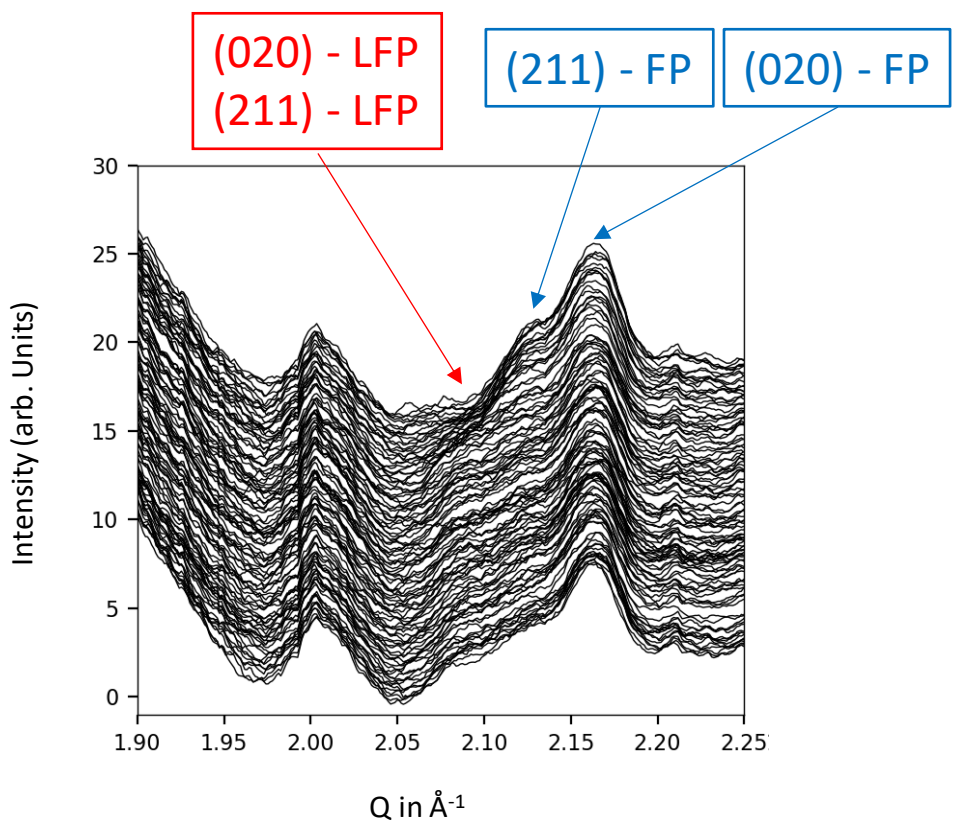

Figure S3. – Evolution of the WAXS pattern at the reference electrode position during the overcharge (going up). LFP peaks located at 2.07 Å disappear while the FP peak gain in intensity at the end of the overcharge.

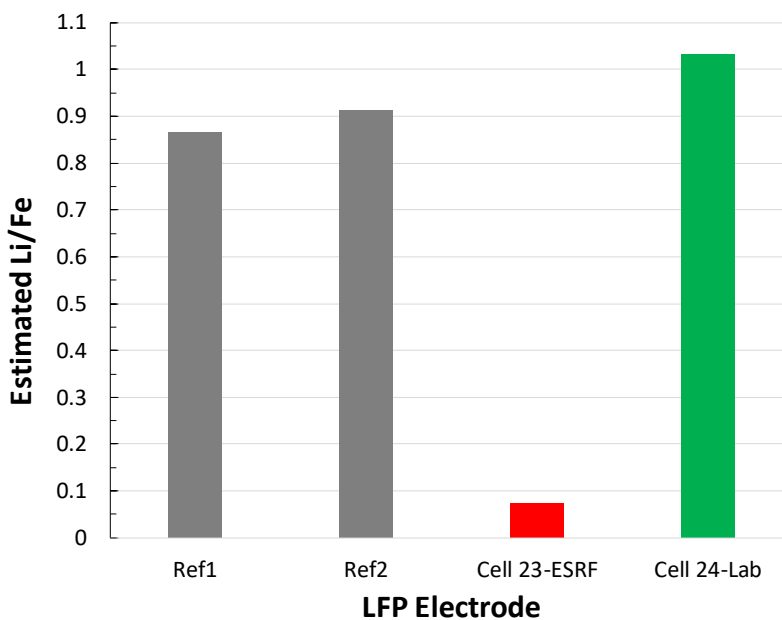

Fig. S4. Estimated Li/Fe content based on the first delithiation capacity over averaged delithiation capacities measured for recovered LFP and pristine LFP reference electrodes measured in coin cells vs. Li at 5 µA. This estimation assumes that the averaged delithiation capacities correspond to 1 Li/Fe. In reality, these capacities were lower than 1 Li/Fe since the current of 5 μA corresponding to about C/5 (capacities around 25 μAh) is insufficiently slow to allow the full (de)lithiation of LFP electrodes. In addition, there can be a contribution from the electrode recovery process from the cell. In turn, this can explain why the Li/Fe is slightly higher than 1 in the case of the Cell24-Lab cell. This figure is for indicative purpose only.

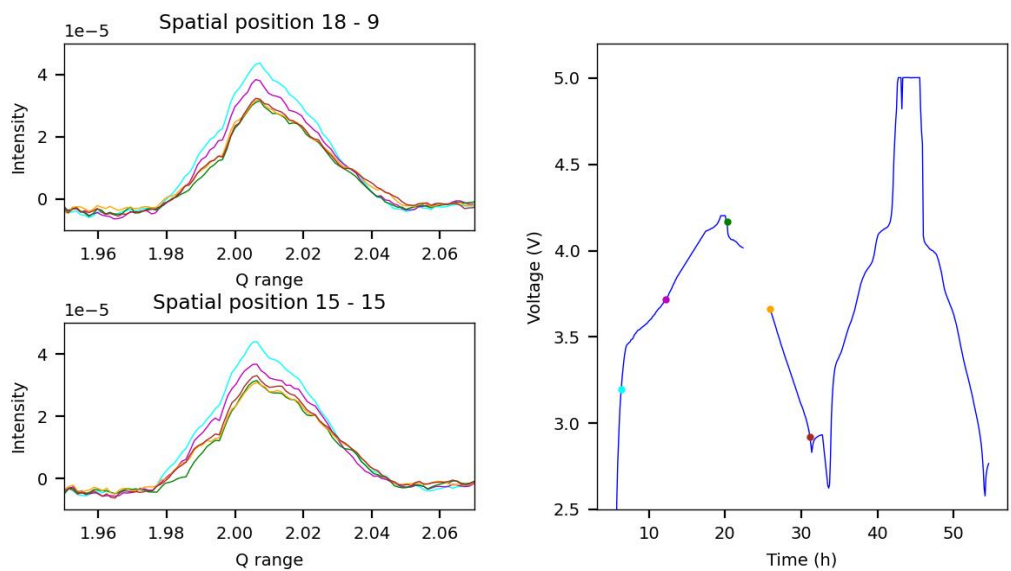

Figure S5: Cristalline silicon peak evolution. Left panel shows diffraction patterns at two different locations in the center of the pouch (pixels (x,y) = (18-9) and (15,15) top and bottom figure). Diffraction patterns are zoomed around 2 Å$^{-1}$ corresponding to the Si (111) peak. Several different patterns are overlaid corresponding to different time during the formation cycle. Right panel shows the electrochemistry with the time at which the diffraction patterns were taken marked as colored dots. Colors of dots and patterns match. It is clear that the intensity of the Si peak decreases during the first charge, indicating amorphisation due to alloying with Li, and stays fairly constant during the subsequent discharge. In fact there is a small intensity increase during discharge (yellow and brown curves) located around 1.985 Å$^{-1}$. We are not sure this is a contribution for crystalline Si as it is close to noise level.

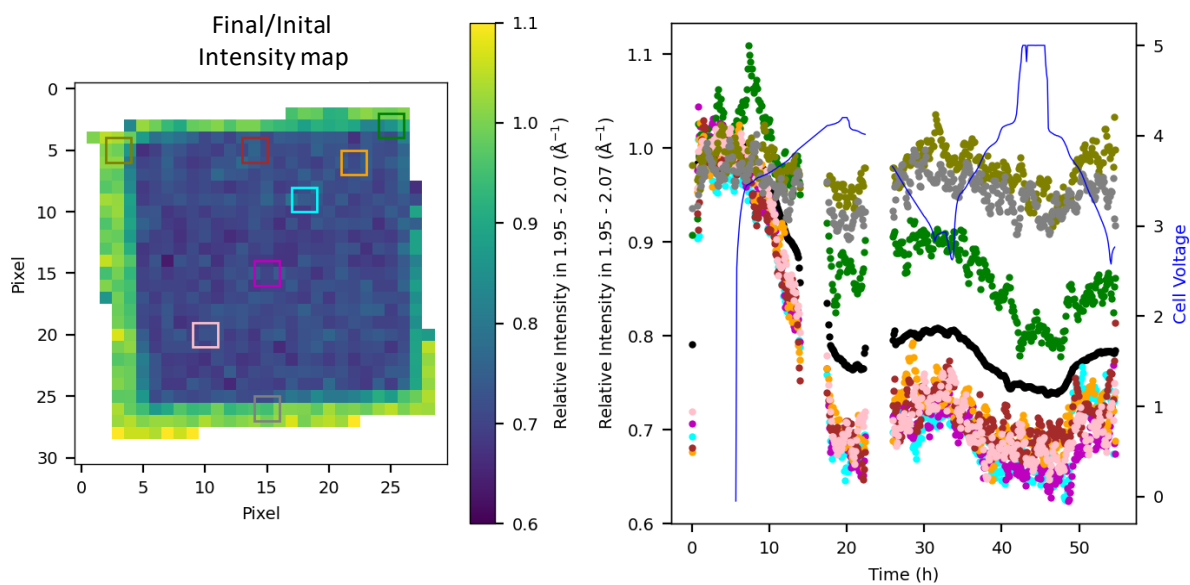

Figure S6: Heterogeneities of silicon lithiation from WAXS data. Left panel is a map of the intensity ratio of the cristalline Si peak before and after the full electrochemical sequence (formation and overcharge). Central regions (corresponding to the electrode stack LNO/Gr-Si) appears blue which means that approx. 70% of the intensity of Si peak is still present after the electrochemical sequence. Outside the stack, the Gr-Si electrode appears green meaning that Si peak intensity did not change much. Right panel shows the intensity ratio as a function of time in average (black) and at specific position in the pouch (colored dots matching with the colored rectangles on the left panel). Blue curve is the electrochemical voltage profile. Overall, this graph clearly shows a continuous decrease of the Si peak intensity during the first charge. The intensity seems to be changing during the subsequent discharge and cycle, but we are unsure if this is due to Si (see figure caption of Figure S5).

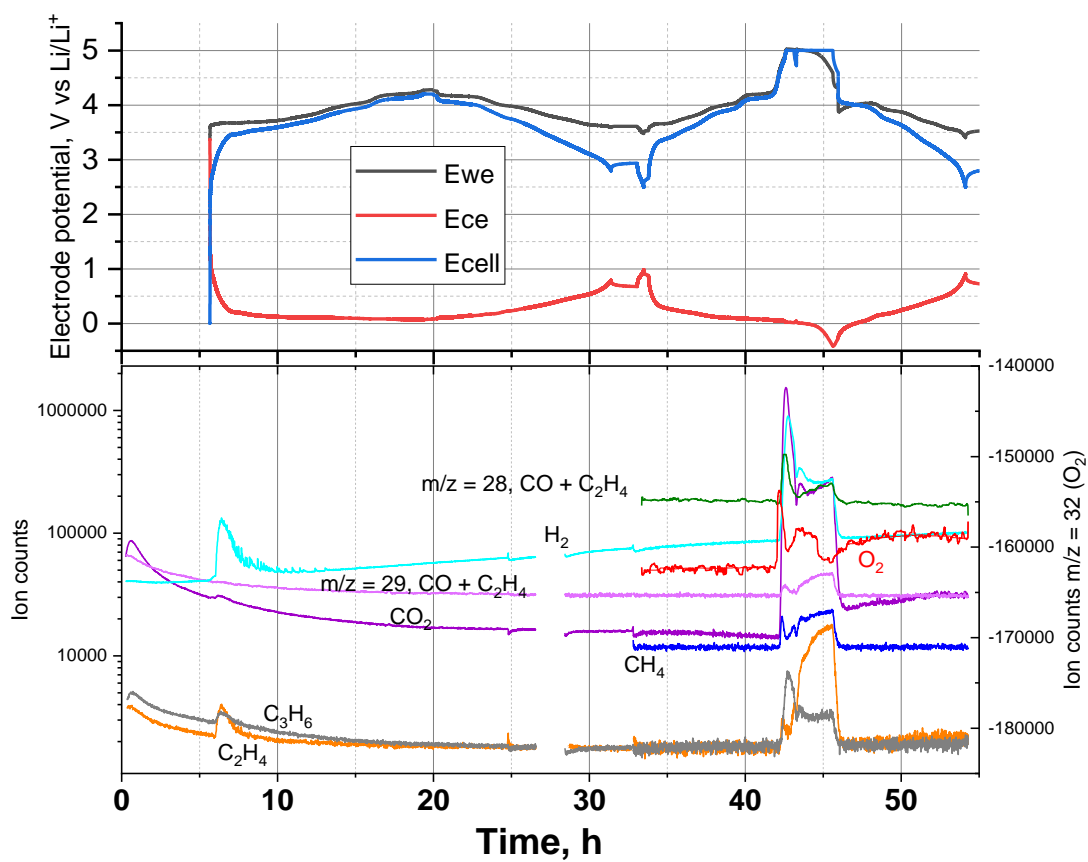

Fig. S7. Voltage profiles for the LNO/Gr-Si pouch cell tested at ESRF during its electrochemical cycling with a cut-off voltages of 4.2V and 5.0V vs. Li/Li$^+$ for the 1$^{st}$ and 2$^{nd}$ cycles, respectively (upper part); corresponding gas evolution curves obtained by OEMS method (lower part).

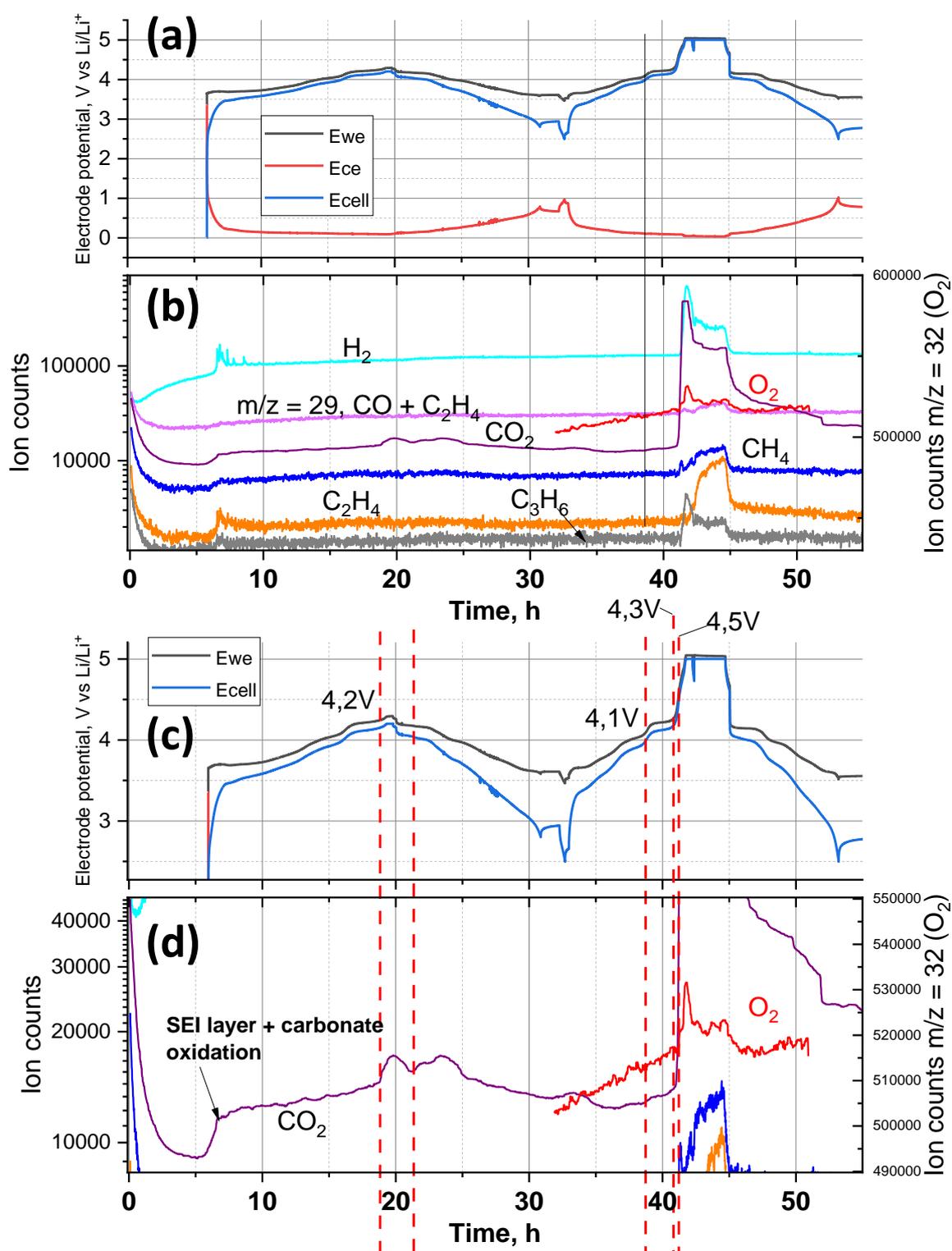

Fig. S8. Voltage profiles for a LNO/Gr-Si pouch cell tested outside the beamline during its electrochemical cycling with a cut-off voltages of 4.2V and 5.0V vs Li/Li$^+$ for the 1$^{st}$ and 2$^{nd}$ cycles, respectively (a) and (c); corresponding gas evolution curves obtained by OEMS method (b) enlargement in $CO_2$ and $O_2$ evolution onset points (d).

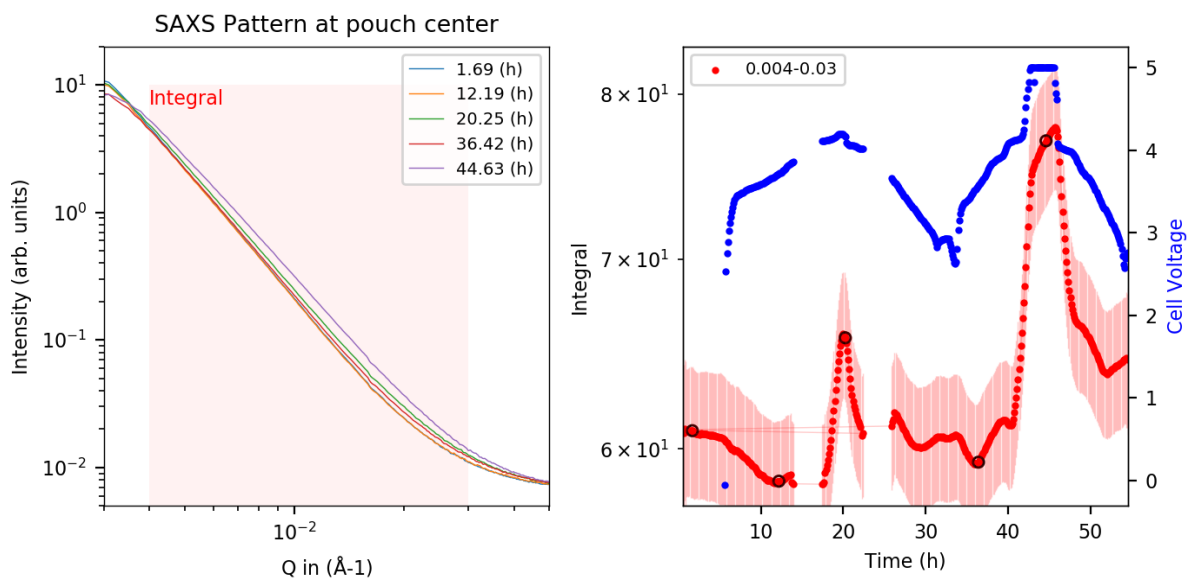

Figure S9: Left panel shows SAXS patterns collected at the center of the pouch cell at different time during the electrochemical sequence (shown by red dots on right panel). Clear changes are observed during the cycling mostly due to intensity increase and possibly small change of slope. Red square corresponds to the region which was integrated and shown Figure 2 and 3. Right panel is the electrochemical profile (blue) together with the integrated SAXS intensity. Error bars correspond to the standard deviation across the pouch cell surface.

**SAXS discussion.**

Exemplary 1D SAXS profiles represented in log-log scale and measured during the electrochemical sequence at several times (averaged from pixels in the center of the cell) are presented in Figure S9. A rather unshaped decaying intensity is measured, which indicate the lack of long-range order and/or the absence of well-defined nanosized domains or structures with sufficient contrast with respect to their neighboring environment. A quantitative analysis of these profiles and their changes depending on the location in the cell is difficult because:

1) The glassy carbon window produces an elevated high-Q background at $Q>3.10^{-2}$ Å$^{-1}$, potentially masking intensity variations arising from modifications in the range of few tens of nms, where expansion and shrinking of the nanosilicon is expected to manifest.

2) The intensity is the sum of small angle scattering arising from all textured materials in the anode and cathode, and from the separator.

1) Absolute integrated SAXS intensity depend on the position in the cell. It is highest at the reference electrode, and lowest for positions only containing Gr-Si electrode. This difference is simply due to the increase of material in the beam path (reference electrode + LNO + Gr-Si compared to Gr-Si only).

However, a qualitative analysis of the differential profiles in time and integration of the signal over specific Q-range can be used to detect specific features evolutions. The Q-range [$4.10^{-3} – 3.10^{-2}$] Å$^{-1}$ probed in this experiment typically relates to structural changes at a scale of 20-150 nms. In a simplified two-phase system where particles (as silicon) are embedded in a medium (the carbon-binder-pores phase wetted with electrolyte, for instance), the SAXS intensity I(Q) would be proportional to the volume fraction of these particles and the square of the contrast term $(\Delta\rho)^2$ that directly depends on the difference in electronic density between the particles and the medium. I(Q) is also modulated by form and structure factors describing the sizes, shapes and organization of the various particles. Typically, an assembly of nanoscale particles may produce a shaped signal presenting local minima, maxima or oscillatory features characteristics of mean dimensions of, or separation distances between, scattering objects. In case of graphite-composite anodes, the pristine SAXS profile is often shapeless due to a lack of contrast, but I(Q) changes develop with lithiation, which can be analysed to obtain indications of nanodomains swelling. Note that graphite lithiation is expected to lead to a decreased I(Q) signal due to decreasing electronic density constrast between $Li_xC_6$ and the medium, when x varies from 0 to 1. The signal is usually following a Porod's law, *e.g.* the intensity varies as $Q^{-4}$ due to the existence of neat interfaces between micrometric graphite particles and the pores. In contrast, the silicon intensity may increase while forming $Li_xSi$ alloys or exhibit a non-monotonous behavior depending on the nature of its surroundings, because of complex contrast terms evolution due to core-shell mechanism, $SiO_2$ lithiation, SEI evolution, etc.). Also, the 300% volume changes expected on lithiation may be revealed by I(Q) shaped by evolving form and structure factors, *e.g.* data might be described with metrics related to typical core-shell sizes, for instance. Nevertheless, the number of parameters to take into account for describing a composite silicon-graphite anode, intrinsically formed of many evolving phases and a hierarchy of scales, is impeding a full analytical treatment of the SAXS data.

In case of large micron-sized objects, the decaying SAXS intensity arises from interfaces and may remain insensitive to particles evolutions in size, shape, and structure. In the pouch cell studied here, the only material organized at the nanoscale is silicon (typically composed of 30 nms sized domains), which is expected to expand and shrink during charge/discharge. In the present conditions, however silicon lithiation is partial (as seen from the limited crystalline Si peak variations) and, moreover, the amount of material in the composite anode is low. Regarding the cathode particles, they are much larger and contribute in the Q-range by a sloped signal due to interfaces detectable at very small Q-values. This intensity can be analysed in principle using the Porod's representation or a decaying $Q^{-\alpha}$ law, where the exponent might relate to the dimensionality of the particles and fractality of the surface.

As seen on Figure S9 right panel, the SAXS integrated intensity features two clear changes located at the end of both charges, especially during the overcharge. Interestingly, this increase is not visible for positions containing only Gr-Si, despite the apparent but moderate lithiation of Gr and Si. Moreover, these features do not seem to be correlated with the presence of gas because very little intensity change is observed during the SEI formation. Hence, we suggest that the increase in SAXS intensity is mostly correlated to microstructural change at the cathode side at high voltage.

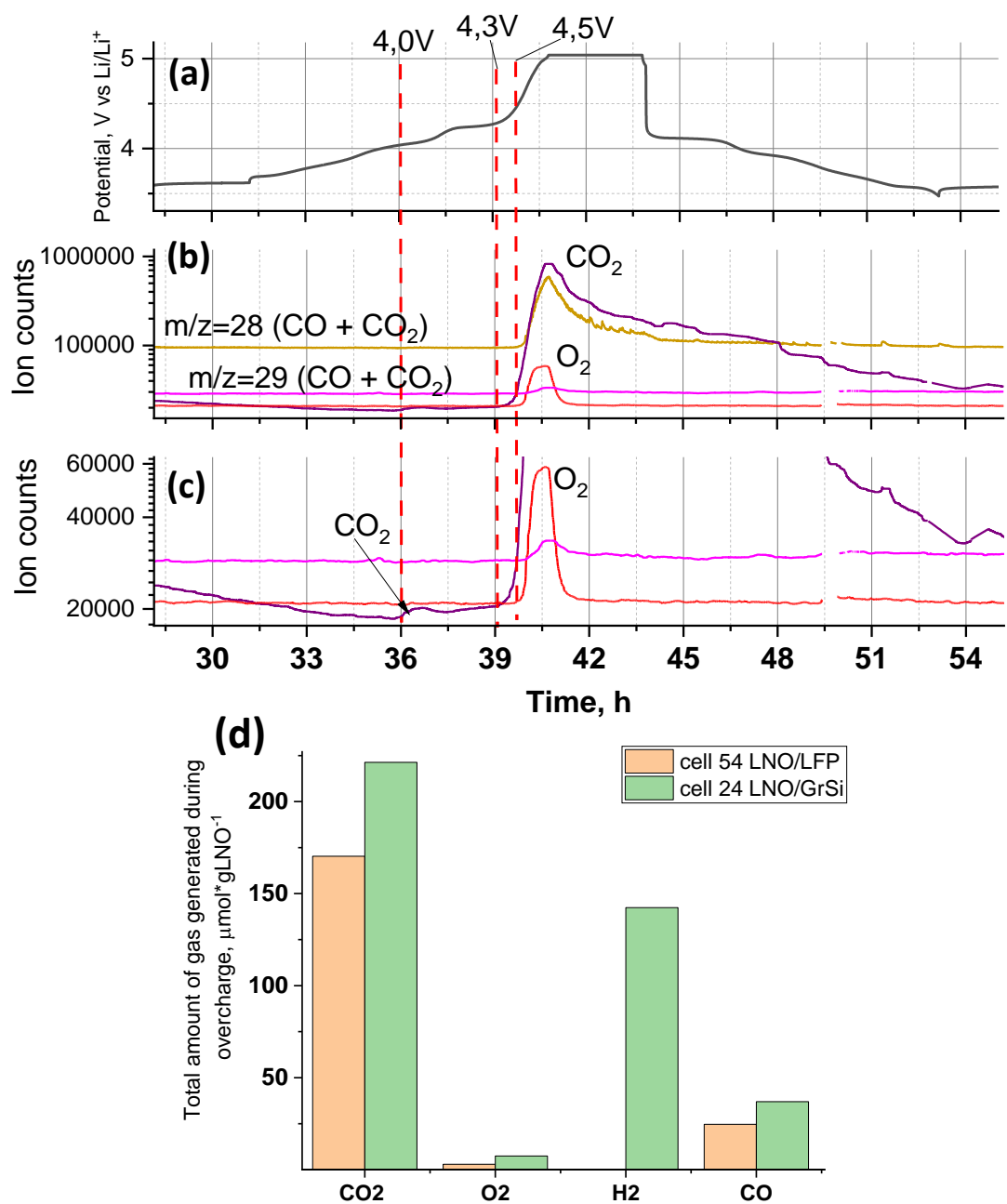

Fig. S10. Voltage profile for the LNO/FP (delithiated LiFePO$_4$) pouch cell tested outside the beamline during its electrochemical cycling with a cut-off voltage of 5.0V vs. Li/Li$^+$ (a); corresponding gas evolution curves obtained by OEMS method (b) and enlargement in CO$_2$ and O$_2$ gas profiles (c). Total amounts of gases evolved during overcharge cycle (d) for the cells containing LNO/delithiated LFP (cell 54) and LNO/GrSi electrodes (cell 24), respectively. Both experiments were conducted in a laboratory and not at the beamline.

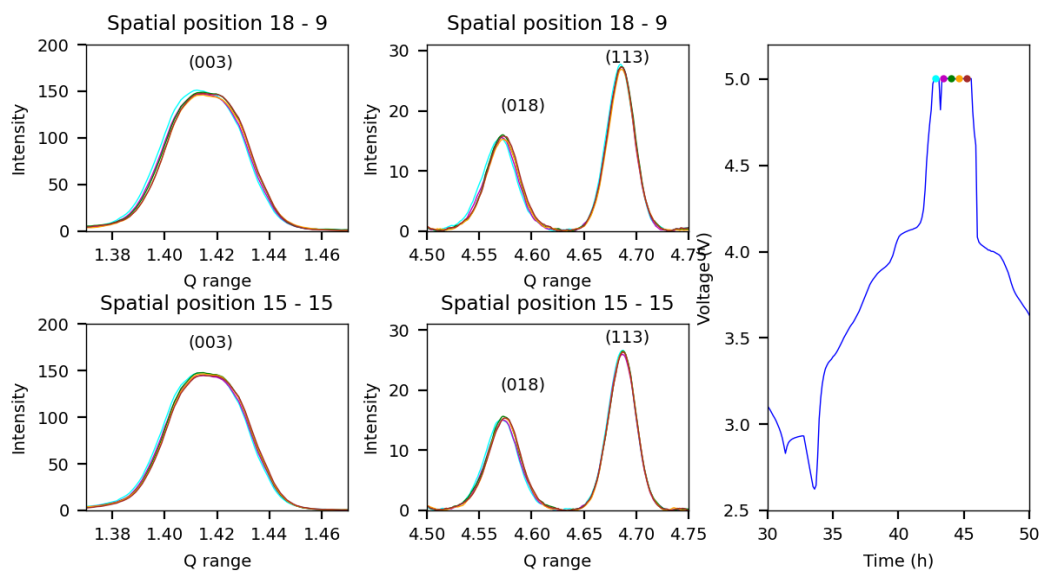

Figure S11: left and middle panels show diffraction patterns at two different locations in the center of the pouch (pixels (x,y) = (18-9) and (15,15) top and bottom figure). Left and mid panel show different Q ranges, namely around the (003) and (018) and (113) reflections, respectively. Several different patterns are overlaid corresponding to different time in the voltage hold. Right panel shows the electrochemistry with the time at which the diffraction patterns were taken marked as colored dots. Colors of dots and patterns match. It is clear that the (018) shift towards high angle during the hold while the (113) shift less. This proves that the *c* lattice parameter of LNO decreases during hold. The (003) also shifts and does not feature any strong asymmetry not supporting the presence of O1 phase or stacking faults.

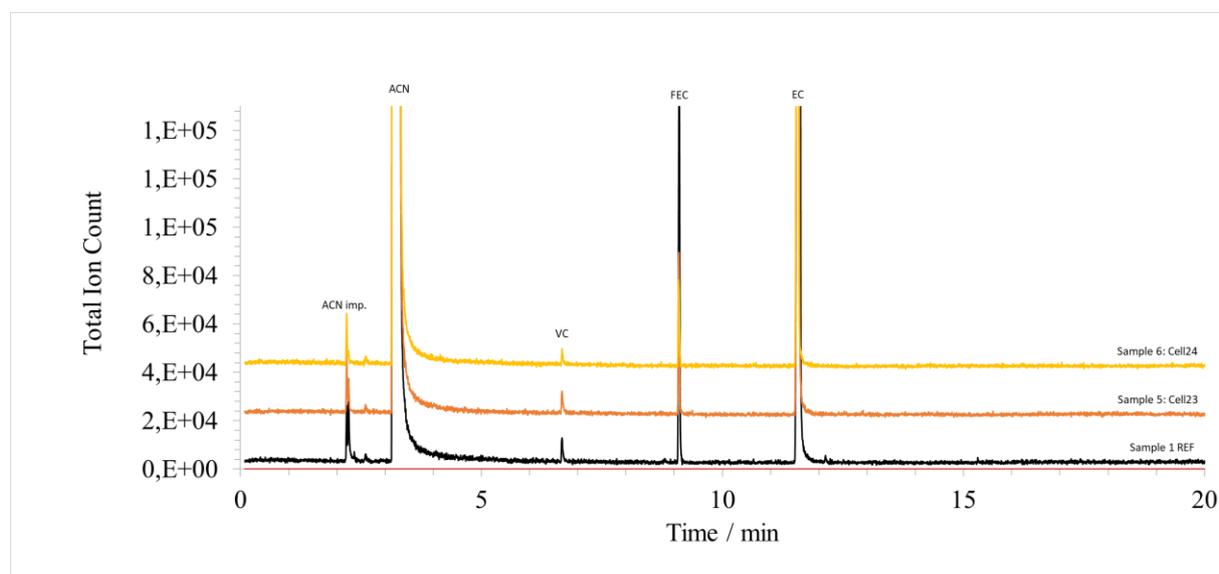

Fig. S12. Gas chromatograms obtained for the reference electrolyte EC/1.3 M LiPF$_6$ with 10 wt.% FEC (REF), for the cell after cycling at ESRF (cell 23) and in the laboratory (cell 24).

The electrolytes were diluted in acetonitrile. VC impurity was identified in the reference electrolyte and its proportion increased after cycling.

Table S1. Results of GC-MS peaks integration

| Sample | Ratio EC/FEC | Ratio EC/VC |
|---|---|---|
| **Reference** | 10,7 | 133,1 |
| **Cell 23 ESRF** | 17,5 | 77,3 |
| **Cell 24 Lab** | 20,7 | 79,7 |

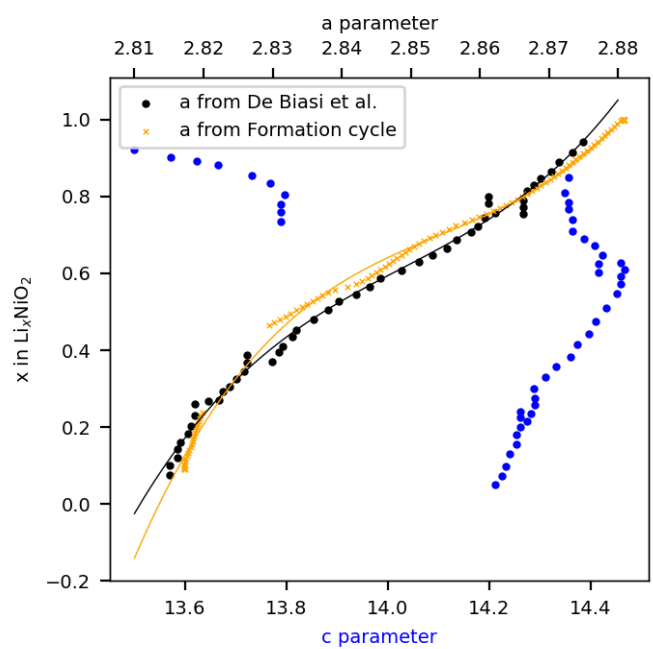

Figure S13: Cell parameter variation of LNO depending on Li concentration (x in $Li_xNiO_2$). Blue dots, black dots and orange crosses are *c* parameter from literature[1], *a* parameter from litterature[1], and *a* parameter determined from (101) and (003) peak position during the charge of the formation cycle performed in this experiment. Orange and black lines are third order polynomial fits, the orange line is the one used in this work to estimate Li concentration from *a* parameter.

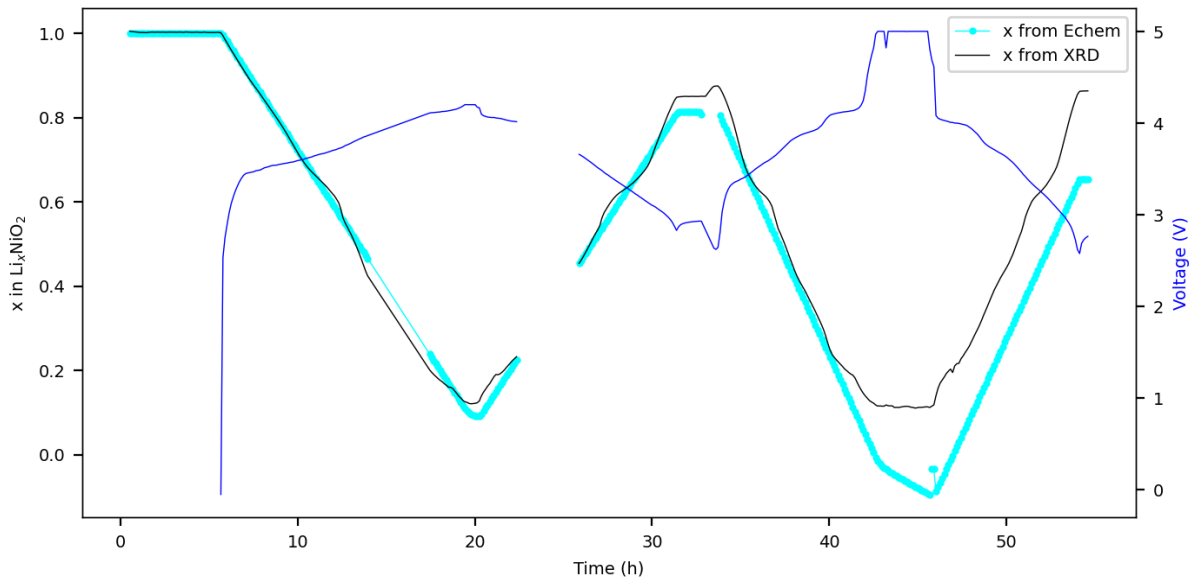

Figure S14: Average x in Li$_x$NiO$_2$ obtained from electrochemistry (just basically converting the capacity) and XRD. Good agreement is observed for the 1$^{st}$ cycle. Deviation is observed during the overcharge since the x estimated from electrochemistry is much lower than the one obtained from XRD. This is likely due to two effects: (1) the most important one is that a substantial part of the current is devoted to electrolyte oxidation and not Li removal from LiNiO$_2$, 2) the *a* parameter vs. Li concentration calibration is not accurate at high voltage because of the lack of low Li concentration reference of Li$_x$NiO$_2$.

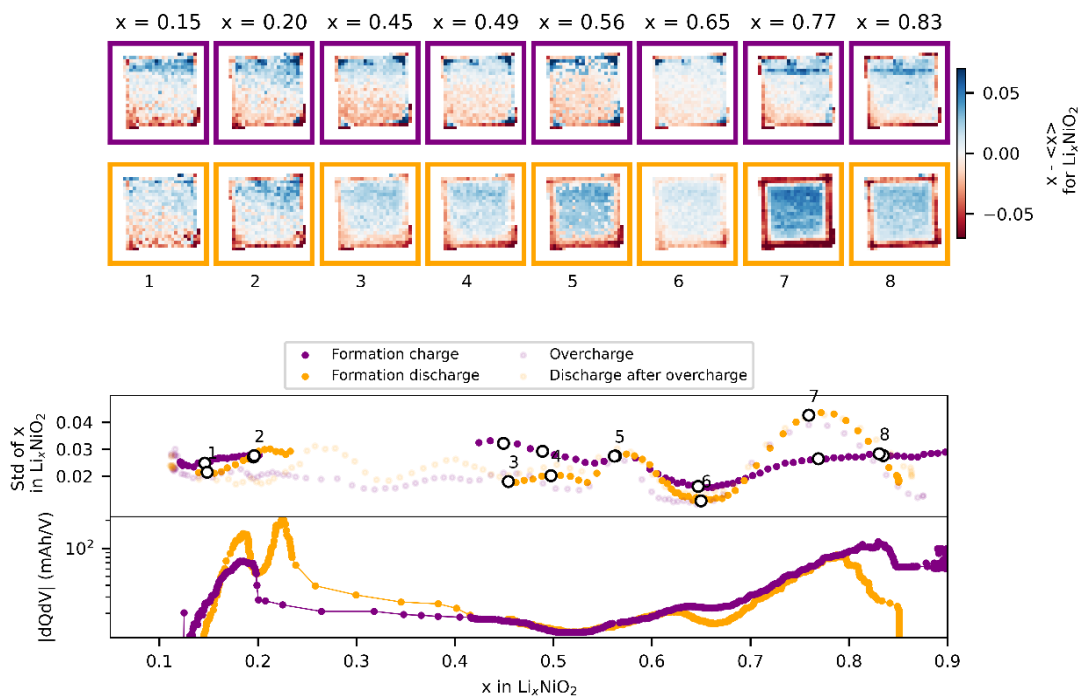

Figure S15: Reaction heterogeneity in LNO during formation cycle. Top panel shows Li concentration deviation maps for the formation cycle charge (top) and discharge (bottom). Li

concentration deviation maps are calculated by subtracting the spatial average across the entire electrode from the local Li concentration resulting in blue and red regions corresponding to zones having more and less Li compared to the average electrode. The bottom panel shows the standard deviation of the Li concentration histogram for each maps – which is a quantitative description of the amplitude of the spatial heterogeneity – together with the dQ/dVdV of the formation cycle calculated for the LNO. Purple colors are used for the charges and orange for the discharges.

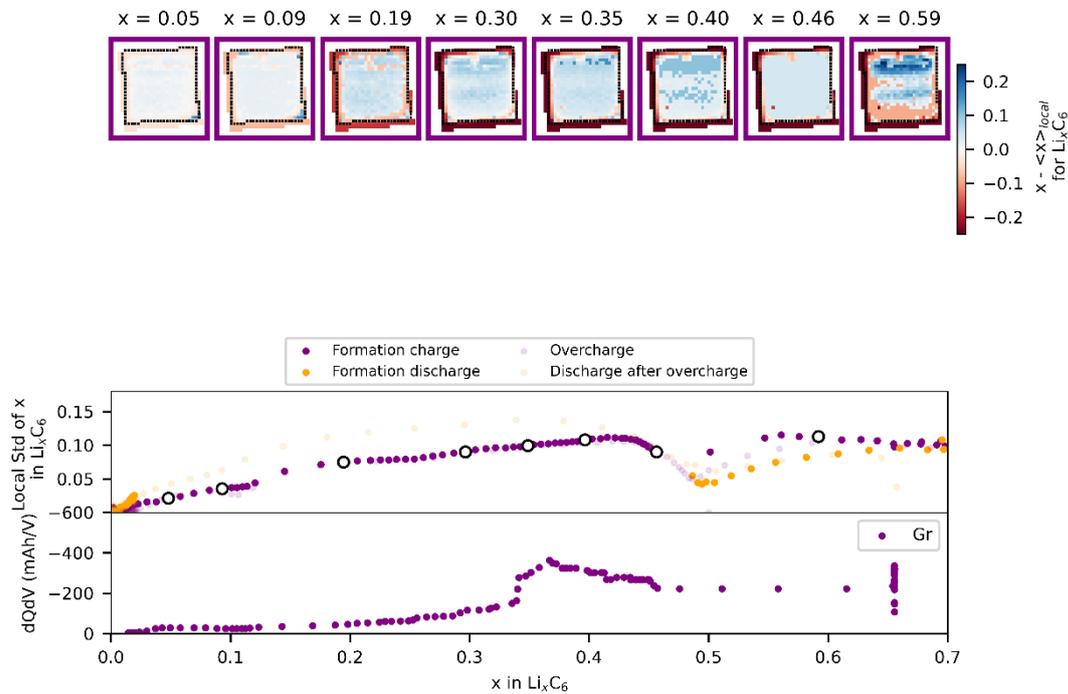

Figure S16: Reaction heterogeneity in graphite during formation cycle charge. Top panel shows Li concentration deviation maps for the charge of the formation cycle. Li concentration deviation maps are calculated as follow : for each pixel, the local Li concentration (of the pixel) is subtracted the average Li concentration in the graphite electrode facing the LNO electrode resulting in blue and red regions corresponding to zones having more and less Li compared to the average electrode. The bottom panel shows the standard deviation of the Li concentration histogram for each maps – which is quantitative description of the amplitude of the spatial heterogeneity – together with the dQ/dVdV of the formation cycle calculated for the graphite-silicon electrodes. Purple colors are used for the charges and orange for the discharges.